\documentclass[english,aps,twocolumn,prd,superscriptaddress]{revtex4-1}
\usepackage[T1]{fontenc}
\usepackage[latin9]{inputenc}
\usepackage{babel}
\usepackage{bm}
\usepackage{amsmath}
\usepackage{amsthm}
\usepackage{amssymb}
\usepackage{graphicx}
\usepackage[colorlinks=true,citecolor=blue,
linkcolor=blue,urlcolor=blue]{hyperref}
\hypersetup{
 colorlinks=true,urlcolor=blue,anchorcolor=blue,citecolor=blue,filecolor=blue,linkcolor=blue,menucolor=blue,pagecolor=blue,linktocpage=true}

\makeatletter
\theoremstyle{plain}

\theoremstyle{remark}

\usepackage{physics}
\usepackage{slashed}
\DeclareMathOperator{\sgn}{sgn}

\makeatother

\providecommand{\remarkname}{Remark}
\providecommand{\theoremname}{Theorem}

\renewcommand{\b}[1]{{\boldsymbol{#1}}}
\renewcommand{\c}[1]{\mathcal{#1}}
\newcommand{\nn}{\nonumber}
\renewcommand{\leq}{\leqslant}
\renewcommand{\geq}{\geqslant}

\begin{document}
\title{Symmetry-breaking effects of instantons in parton gauge theories}
\author{G. Shankar}
\email{sankaran@ualberta.ca}
\affiliation{Department of Physics, University of Alberta, Edmonton, Alberta T6G 2E1, Canada}
\author{Joseph Maciejko}
\email{maciejko@ualberta.ca}
\affiliation{Department of Physics, University of Alberta, Edmonton, Alberta T6G 2E1, Canada}
\affiliation{Theoretical Physics Institute, University of Alberta, Edmonton, Alberta T6G 2E1, Canada}
\date{\today}
\begin{abstract}
Compact quantum electrodynamics (CQED$_3$) with Dirac fermionic matter provides an adequate framework for elucidating the universal low-energy physics of a wide variety of (2+1)D strongly correlated systems. Fractionalized states of matter correspond to its deconfined phases, where the gauge field is effectively noncompact, while conventional broken-symmetry phases are associated with confinement triggered by the proliferation of monopole-instantons. While much attention has been devoted lately to the symmetry classification of monopole operators in massless CQED$_3$ and related 3D conformal field theories, explicit derivations of instanton dynamics in parton gauge theories with fermions have been lacking. In this work, we use semiclassical methods analogous to those used by 't~Hooft in the solution of the $U(1)$ problem in 4D quantum chromodynamics (QCD) to explicitly demonstrate the symmetry-breaking effect of instantons in CQED$_3$ with massive fermions, motivated by a fermionic parton description of hard-core bosons on a lattice. By contrast with the massless case studied by Marston, we find that massive fermions possess Euclidean zero modes exponentially localized to the center of the instanton. Such Euclidean zero modes produce in turn an effective four-fermion interaction---known as the 't~Hooft vertex in QCD---which naturally leads to two possible superfluid phases for the original microscopic bosons: a conventional single-particle condensate or an exotic boson pair condensate without single-particle condensation.
\end{abstract}
\maketitle

\section{Introduction}

The parton or projective construction is one of the most versatile and conceptually fruitful approaches to a theoretical understanding of strongly correlated systems~\cite{wen2004}. This approach is based on rewriting microscopic degrees of freedom in terms of fractionalized ones that are charged under an emergent gauge field, and thus transform projectively under microscopic symmetries. The emergent gauge structure strongly constrains the low-energy physics, which is progressively revealed as high-energy degrees of freedom are integrated out. A lattice gauge theory with dynamical gauge fields first emerges, and is then replaced by a continuum gauge theory once lattice-scale fluctuations have been decimated. The universal low-energy physics of the original quantum many-body system is then dictated by the infrared fate of this continuum parton gauge theory.

Fractionalized phases of matter, such as spin liquids, fractionalized Fermi liquids, or fractional quantum Hall states, correspond to deconfined phases of parton gauge theories. Whether such phases exist at all for $U(1)$ parton gauge theories in 2+1 dimensions---our prime focus---is a nontrivial question, due to the strong infrared relevance of the gauge coupling and the ensuing tendency to confinement. Nonperturbative confinement-inducing effects in such gauge theories, notably monopole-instantons~\cite{polyakov1975,polyakov1977,polyakov1987}, can be suppressed by a variety of mechanisms, including large-flavor screening effects~\cite{appelquist1986,hermele2004}, the Higgs mechanism~\cite{wen1991}, and Chern-Simons topological masses~\cite{pisarski1986,affleck1989}. If the suppression of monopole-instantons does obtain, the appropriate fractionalized phase is adiabatically connected to a weakly coupled phase of the parton gauge theory, despite being highly nonperturbative from the point of view of the microscopic Hamiltonian.

While fractionalized phases are thus perturbatively accessible in the parton framework, conventional broken-symmetry phases are more difficult to describe, as nonperturbative confinement effects must then necessarily play a role. An ability to describe conventional phases within the framework of parton gauge theory is however necessary for overall consistency of the theory, as well as to understand the mechanism underlying confinement transitions between a fractionalized phase and proximate conventional phases. This question was studied carefully in recent work~\cite{song2019,song2020} in the context of the Dirac spin liquid, described at low energies by $U(1)$ quantum electrodynamics (QED$_3$) with four flavors of two-component massless Dirac fermions~\cite{affleck1988,kim1999,rantner2002,hermele2005,*hermele2007}. Extending earlier work by Alicea and collaborators~\cite{alicea2005,alicea2005b,alicea2006,alicea2008}, Song {\it et al.}~\cite{song2019,song2020} utilized the state-operator correspondence of conformal field theory~\cite{borokhov2002} to determine the quantum numbers of monopole operators $\c{M}$ for microscopic realizations of the Dirac spin liquid state on various lattices. The insertion of a (single) monopole operator in the Hamiltonian formalism corresponds to an instanton event in (2+1)D spacetime whereby a localized source of $2\pi$ magnetic flux is suddenly added to the system~\cite{polyakov1975,polyakov1977,polyakov1987}. In the Hamiltonian picture, a conventional phase is argued to be accompanied by a monopole condensate $\langle\c{M}\rangle\!\neq\!0$ which confines excitations with nonzero gauge charge, gives a mass to the emergent photon, and breaks physical symmetries if $\c{M}$ transforms nontrivially under the latter.

While these arguments are undoubtedly correct, there exist few explicit computations of the nonperturbative {\it dynamics} that would substantiate these general symmetry considerations. Song {\it et al.} assume a two-step scenario in which a gauge-invariant fermion mass bilinear first acquires an expectation value, a process described by an effective theory of the QED$_3$-Gross-Neveu-Yukawa type~\cite{janssen2017,ihrig2018,zerf2018,
dupuis2019,zerf2019,boyack2019,boyack2019b,zerf2020,janssen2020,boyack2021,xu2019,wang2019} in which compactness of the gauge field is assumed to not play a role. After the fermionic matter is gapped out, instanton proliferation is further assumed to proceed as in the pure compact gauge theory~\cite{polyakov1975,polyakov1977,polyakov1987}.

In the presence of fermionic matter, however, gauge instantons may be accompanied by fermion zero modes (ZMs)~\cite{rubakov}, which can qualitatively affect the dynamics of instanton proliferation. Such Euclidean ZMs are traditionally associated with massless fermions, and are responsible for symmetry-breaking effects in the fermion sector. In (3+1)D Yang-Mills theory with massless fermions in the fundamental representation, 't Hooft showed~\cite{thooft1976,thooft1976a,thooft1986} that fermion ZMs on the Belavin-Polyakov-Schwartz-Tyupkin instanton~\cite{belavin1975} are responsible for the explicit breaking of chiral symmetry in the fermion sector, in a manner consistent with the Adler-Bell-Jackiw anomaly equation~\cite{adler1969,bell1969}. Fermion ZMs on gauge instantons in the (2+1)D Georgi-Glashow model~\cite{polyakov1977,polyakov1987} with massless fermions in the adjoint representation were shown by Affleck, Harvey, and Witten~\cite{affleck1982} to possibly lead to spontaneous breaking of the global $U(1)$ fermion number conservation symmetry. In both cases, Euclidean fermion ZMs generate, via resummation of the semiclassical instanton gas, an effective fermionic interaction---the 't Hooft vertex---that manifests the desired broken symmetry. The existence of fermion ZMs in the above theories is guaranteed by the Atiyah-Singer index theorem in (3+1)D~\cite{atiyah1963} and the Callias index theorem in (2+1)D~\cite{callias1978,bott1978}. The latter in particular crucially relies on the non-Abelian nature of the gauge field and the presence of a scalar Higgs field in the adjoint representation which winds nontrivially at infinity in the instanton solution~\cite{polyakov1977}.

The examples above involve non-Abelian gauge fields and do not directly apply to our prime focus, but nonetheless suggest that fermion ZMs on gauge instantons may play an important role in the description of conventional phases and their broken symmetries in $U(1)$ parton gauge theories. A natural starting point to investigate this question is compact QED$_3$ with massless Dirac fermions, relevant for the Dirac spin liquid. At late times and long distances, the Polyakov lattice instanton can be modeled as a Dirac monopole in 3D Euclidean space. The corresponding Euclidean massless Dirac equation was studied by Marston~\cite{marston1990}, but shown by explicit calculation to {\it not} exhibit any normalizable ZM bound to the instanton. Further, there appears to exist no generalization of the Callias index theorem to compact QED$_3$~\cite{unsal2008}, despite the similar infrared fate of the (2+1)D Georgi-Glashow model and compact QED$_3$ without fermions~\cite{polyakov1975,polyakov1977,polyakov1987}. In the absence of explicit fermion ZM solutions or a general theorem guaranteeing their existence, their relevance to the infrared dynamics of $U(1)$ parton gauge theories is at best speculative.

We emphasize here that we are interested in fermion ZMs bound to instantons in noncompact Euclidean spacetime $\mathbb{R}^3$, as opposed to ZMs of the Dirac Hamiltonian on a 2-sphere $S^2$ surrounding a monopole insertion in the state-operator correspondence of conformal QED$_3$~\cite{borokhov2002}. The existence of the latter ZMs is guaranteed by the Atiyah-Singer index theorem applied to the massless Dirac operator on the compact space $S^2$. As the Marston calculation~\cite{marston1990} indicates, however, the existence of Hamiltonian ZMs in the latter context does not automatically imply the existence of Euclidean ZMs in noncompact spacetime.

In this paper, we present a study of nonperturbative effects in a $U(1)$ parton gauge theory, in which we show by explicit calculation that Euclidean fermion ZMs bound to gauge instantons exist and lead to symmetry-breaking effects. The gauge theory we consider arises as the effective continuum description of interacting lattice bosons in the vicinity of a multicritical point separating superfluid, Mott insulating, and fractional quantum Hall ground states~\cite{barkeshli2014}. While the parton description is introduced as a means to access the fractional quantum Hall state, in which a Chern-Simons term for the emergent $U(1)$ gauge field leads to deconfinement, our focus here is on the nonperturbative gauge dynamics that obtains in the superfluid phase, which must result simultaneously in the confinement of gauge-charged excitations and the spontaneous breakdown of the global $U(1)$ boson number conservation symmetry. Ref.~\cite{barkeshli2014} argues from general considerations that the Affleck-Harvey-Witten mechanism~\cite{affleck1982} should be operative and yield the desired physics, but does not provide an explicit derivation of the underlying instanton dynamics. Here we show by explicit calculation that, by contrast with massless QED$_3$, QED$_3$ with {\it massive} Dirac fermions admits normalizable Euclidean ZM solutions in a $U(1)$ instanton background. Such solutions are exponentially localized to the center of the instanton with a length scale inversely proportional to the fermion mass. Using semiclassical methods~\cite{thooft1976,thooft1976a,thooft1986}, we then explicitly compute the 't Hooft vertex induced by those ZMs and show that it naturally leads to {\it two} possible superfluid phases: a conventional superfluid phase with single-particle condensation~\cite{barkeshli2014}, but also an exotic paired superfluid phase with a residual $\mathbb{Z}_2$ symmetry.

The rest of the paper is structured as follows. We briefly review Ref.~\cite{barkeshli2014}'s parton description of the interacting boson problem in Sec.~\ref{sec:LGT}. In Sec.~\ref{sec:thetvac}, we formulate the imaginary-time partition function of the system in a way that makes the contribution of Polyakov instantons manifest, and allows us to introduce $\theta$ parameters analogous to those of 4D Yang-Mills theory~\cite{jackiw1976,callan1976}. In Sec.~\ref{sec:massmars}, we show that massive fermions support Euclidean zero modes localized on such instantons, and discuss their relationship to Hamiltonian (quasi-)zero modes in canonical quantization. Sec.~\ref{sec:hooft} details the calculation of the 't Hooft vertex and Sec.~\ref{sec:more} explores its symmetry-breaking consequences. We end the paper with brief concluding remarks in Sec.~\ref{sec:concl}; accessory technical results are collated in Appendices~\ref{app:monmisc} and \ref{app:SAop}.


\section{Parton gauge theory}\label{sec:LGT}

We begin by reviewing the parton gauge theory introduced in Ref.~\cite{barkeshli2014}. We consider a system of charge $+1$ (in appropriate units) hard-core
bosons on a 2D lattice described by operators $b(\bm{x})$
and $b^{\dagger}(\bm{x})$. The hard-core condition imposes on these
operators the algebra 
\begin{align}
[b(\bm{x}),b^{\dagger}(\bm{x}')] & =[1-2b^{\dagger}(\bm{x})b(\bm{x})]\delta_{\bm{xx}'},\\{}
[b(\bm{x}),b(\bm{x}')] & =[b^{\dagger}(\bm{x}),b^{\dagger}(\bm{x}')]=0.
\end{align}
The hard-core boson then admits a parton decomposition
\begin{equation}
b(\bm{x})=f_{1}(\bm{x})f_{2}(\bm{x}),\label{eq:part}
\end{equation}
where $f_{1}(\bm{x})$ and $f_{2}(\bm{x})$ are fermion annihilation
operators. We associate the physical boson charge with $f_{1}$ and
couple this to a background gauge field $A_{\mu}$ when it is necessary
to keep track of the physical $U(1)$ symmetry associated with conservation of the boson number. The parton decomposition
(\ref{eq:part}) also introduces a local $SU(2)$ gauge redundancy
$f_{i}(\bm{x})\!\to\!W_{ij}(\bm{x})f_{j}(\bm{x})$, under which the
boson operators remain invariant. In the parton approach~\cite{wen2004}, one first
ignores this gauge structure and postulates a mean-field ansatz for
the partons. Gauge fluctuations above the mean-field fermion ground state are then reintroduced, which ensures the parton dynamics is projected onto the physical boson Hilbert space. In what follows, we shall assume a mean-field
ansatz for the partons that breaks the $SU(2)$ redundancy down to
a $U(1)$ subgroup, for example via a lattice analog of the Higgs mechanism~\cite{wen1991}, which leaves a single gauge boson massless. Under the leftover $U(1)$
gauge redundancy, the partons $f_{1}$ and $f_{2}$ are assigned gauge
charges $\pm1$ respectively, so that the boson operator remains gauge
invariant.

We further consider a mean-field ansatz for the partons in which $f_{1}$
and $f_{2}$ form independent Chern insulators with Chern numbers
$\pm1$, respectively, described for instance by Haldane models~\citep{haldane1988} or their analog on the lattice of interest. In the vicinity of Chern-number-changing transitions in the parton bandstructure, this theory is described in the continuum limit by a 3D Euclidean Lagrangian, 
\begin{equation}
\mathcal{L}=\sum_{\alpha=\pm}\left[\bar{\psi}_{1\alpha}(\slashed{\partial}-i\slashed{A}+m)\psi_{1\alpha}+\bar{\psi}_{2\alpha}(\slashed{\partial}-m)\psi_{2\alpha}\right],
\end{equation}
where $A_{\mu}$ is the background field that tracks the physical
$U(1)$ symmetry, and $\{\psi_{1\pm},\psi_{2\pm}\}$ are two-component
Dirac fermions obtained in a linearization of the partons $\{f_{1},f_{2}\}$
at the two Dirac points $K_{\pm}$ that generically appear in the parton bandstructure~\footnote{In our convention, the 3D Euclidean Dirac matrices are just Pauli
matrices with $\hat{z}$ being the Euclideanized time direction. Matter
of general charge $e$ gauge transforms as $\psi\!\to\!\psi e^{ie\lambda(x)}$,
and the gauge covariant derivative is $(\partial_{\mu}\!-\!iea_{\mu})$.}. Importantly, the fermion masses for $\psi_{1\pm}$ and $\psi_{2\pm}$ are opposite in sign, since the Chern numbers are opposite in sign for $f_1$ and $f_2$.

For the above mean-field parton ansatz to correspond in fact to a physical state of bosons, we must reintroduce gauge fluctuations. To study the effect of those fluctuations, the lattice fermions (i.e.,
the partons) are minimally coupled to an emergent $U(1)$ gauge field $a_{\mu}$.
For example, the parton $f_{1}$ of gauge charge $+1$ minimally couples
to the gauge field on link $(\bm{x},i)$ as $f_{1\bm{x}}^{\dagger}t_{\bm{x},i}\exp(-ia_{\bm{x},i})f_{1\bm{x}+i}$,
where $t_{\bm{x},i}$ is a hopping integral, $\bm{x}$ is a lattice
site, and $i$ is a lattice vector. The invariance of such a term
under $2n\pi$ shifts of $a_{\bm{x},i}$ can be viewed as a gauge redundancy
or as a true local symmetry. These two perspectives will be discussed
in Sec.~\ref{sec:thetvac}. In either case, at low energies, the
renormalization group endows the emergent field $a$ with dynamics
that preserves this periodicity, implying an effective gauge field
Hamiltonian of the form 
\begin{equation}
H_{g}=\frac{1}{2}\sum_{l}e_{l}^{2}+K\sum_{\square}(1-\cos f_{\square}),\label{eq:Hg}
\end{equation}
where $e_{l}$ is the electric field on link $l$ satisfying $[a_{l},e_{l'}]=i\delta_{ll'}$,
and $f_{\square}$ is the flux (lattice curl) of $a$ through the plaquette
$\square$ (we shall henceforth assume a square lattice for simplicity). The physical Hilbert space of the gauge theory (with fermions)
is the gauge invariant subspace specified by a Gauss constraint. Weak
fluctuations of $a_{\mu}$ correspond to the $K\!\gg\!1$ limit, in
which $H_{g}$ is energetically appeased by $f_{\square}\!=\!2\pi n_{\square}$,
where $n_{\square}\!\in\!\mathbb{Z}$ is a plaquette-dependent integer.
Expanding about any one of these minima leads to the usual Maxwell
theory with a massless photon. However, it is well known that tunneling
events $f_{\square}\!\to\!f_{\square}\!+\!2\pi Q$, where $Q\!\in\!\mathbb{Z}$, on a
plaquette cannot be ignored, for these give the photon a mass exponentially
small in the gauge coupling $K$. These tunneling events, corresponding
to $2\pi Q$ flux insertions on a plaquette, feature as instantons
(Dirac monopoles of charge $Q$) with finite action in the 3D Euclidean
theory \citep{polyakov1987,polyakov1975,polyakov1977}.

In a naïve continuum limit, the effective parton Lagrangian with gauge fluctuations is
\begin{multline}
\mathcal{L}=\sum_{\alpha=\pm}\left[\bar{\psi}_{1\alpha}(\slashed{\partial}-i\slashed{A}-i\slashed{a}+m)\psi_{1\alpha}\right.\\
\left.+\bar{\psi}_{2\alpha}(\slashed{\partial}+i\slashed{a}-m)\psi_{2\alpha}\right]+\frac{1}{4e^{2}}f^{2},\label{eq:ptnL}
\end{multline}
where $e$ is the renormalized gauge coupling\textbf{ }(some function
of the lattice coupling $K$). However, a finite UV regulator (lattice
constant) and the fact that the lattice theory is invariant under
$a\!\to\!a\!+\!2n\pi$ imply that the effects of instantons must be
accounted for in this continuum limit. This theory is termed compact
QED$_3$ (CQED$_3$). We note however that by contrast with the CQED$_3$ theory of the Dirac spin liquid, which also has four flavors of two-component Dirac fermions, the fermions in our case (i) are massive, and (ii) do not all have the same sign of the gauge charge.

\section{$\theta$ parameters and instantons}\label{sec:thetvac}

In this section, we use canonical quantization to derive a path integral representation of the partition function of the pure gauge theory without matter, which makes the contribution of instantons explicit and allows us to introduce $\theta$ parameters~\cite{vergeles1979,brown1997} analogous to those of 4D Yang-Mills theory~\cite{jackiw1976,callan1976}. This sets the stage for our computation of the 't Hooft vertex using path integral methods in Sec.~\ref{sec:hooft}, after explicit fermion ZM solutions in the background of a single instanton are obtained in Sec.~\ref{sec:massmars}.

We begin with the pure gauge theory, described by the Hamiltonian (\ref{eq:Hg}),
which we shall consider in the absence of background charges. This means that the Gauss
constraint on every site is $(\mathrm{div}e)_{\bm{x}}\ket{\Psi}\!=\!0$
on all physical states $\ket{\Psi}$. As stated previously, the invariance
of $H_{g}$ under $2\pi Q$ translations of the gauge flux on a plaquette
can be viewed as either a true local symmetry, or as a gauge redundancy
due to rotor-valued link variables $a_{\bm{x},i}\!\in\!U(1)\!\cong\mathbb{R}/2\pi\mathbb{Z}$.
The former view will be called \emph{minimal compactness}, and the
latter \emph{forced compactness}. In what follows, we shall mostly
be concerned with the ``magnetic limit'' $K\!\gg\!1$, in which gauge
fluctuations are weak.

\subsection{Minimal compactness}\label{sec:minimal}

In the minimal compactness picture, the gauge field $a_{\bm{x},i}\!\in\!\mathbb{R}$.
A general state in the Hilbert space is given by a wavefunctional
\begin{equation}
\Psi[a_{\bm{x},i}]=\braket{\{a_{\bm{x},i}\}}{\Psi},
\end{equation}
where $\{a_{\bm{x},i}\}$ denotes the collection of $a$ on all links,
and $\ket{\{a_{\bm{x},i}\}}$ forms a basis. The electric fields $e_{\bm{x},i}$
generate translations of these wavefunctionals. On a single link,
\begin{equation}
e^{-i\alpha e}\ket{a}=\ket{a+\alpha},
\end{equation}
 which means 
\begin{align}
e^{-i\alpha\frac{\delta}{\delta a}}\Psi[a] & =\bra{a}e^{-i\alpha e}\ket{\Psi}\nonumber \\
 & =\braket{a-\alpha}{\Psi}\nonumber \\
 & =\Psi[a-\alpha].
\end{align}
A gauge transformation $\exp[-i\phi(\mathrm{div}e)_{\bm{x}}]$ on
a site $\bm{x}$ is a translation that leaves all plaquette fluxes
$f_\square$ invariant. Since the magnetic term is a periodic function of
$f_\square$, $H_{g}$ is not only invariant under these gauge transformations,
but also under a discrete group of local flux translations $f_\square\!\to\!f_\square\!+\!2\pi Q$, $Q\!\in\!\mathbb{Z}$ on a plaquette. This group is generated by monopole operators $\mathcal{M}_{Q}(\bar{\bm{x}})$,
where $\bar{\bm{x}}$ denotes a plaquette (or equivalently, a site on the dual lattice),
and 
\begin{equation}
\mathcal{M}_{Q}^{\dagger}(\bar{\bm{x}})f_{\bar{\bm{x}}}\mathcal{M}_{Q}(\bar{\bm{x}})=f_{\bar{\bm{x}}}+2\pi Q.\label{eq:monop}
\end{equation}
This translation is also generated by electric fields, but one must
use an infinite string of fields~\citep{drell1979}, since only the
flux in plaquette $\bar{\bm{x}}$ must be changed. One possibility
is to consider an infinite product of all horizontal links below $\bar{\bm{x}}$,
and non-uniquely define 
\begin{equation}
\mathcal{M}_{Q}(\bar{\bm{x}})=e^{i2\pi Q\sigma_{\bar{\bm{x}}}},\qquad\sigma_{\bar{\bm{x}}}\equiv\sum_{p=-\infty}^{0}e_{\bm{x}+p\hat{x}_{2},\hat{x}_{1}},\label{eq:monop1}
\end{equation}
where $\hat{x}_1,\hat{x}_2$ are unit vectors in the positive horizontal and vertical directions, respectively.

The minimally compact theory has similarities with the Bloch problem
of electrons in a crystal lattice, in which a discrete translation
by a lattice constant is a physical symmetry as opposed to a gauge redundancy.
In the Bloch problem, there occur instantons that tunnel between the
minima of the crystal potential, and the true ground state is a superposition
of all local minima. In minimally compact CQED$_3$, the analogs are monopole-instantons
that tunnel between physically distinct minima $f_{\bar{\bm{x}}}\!=\!2n\pi$
to $f'_{\bar{\bm{x}}}\!=\!2m\pi$ on a given plaquette $\bar{\bm{x}}$.
Since $[H_{g},\mathcal{M}_{Q}(\bar{\bm{x}})]\!=\!0$ on every plaquette,
the physical eigenstates of $H_{g}$ fall in representations of these
symmetries. The irreducible representations of this Abelian group
are all one-dimensional, and are simply phase factors (Bloch theorem).
The eigenstates of $H_{g}$ must thus obey,
\begin{equation}
\forall\bar{\bm{x}},\quad\mathcal{M}_{Q}(\bar{\bm{x}})\ket{\Psi_{n,\theta}}=e^{iQ\theta_{\bar{\bm{x}}}}\ket{\Psi_{n,\theta}},\quad\theta_{\bar{\bm{x}}}\!\in\![0,2\pi),\label{eq:bloch}
\end{equation}
where $\theta_{\bar{\bm{x}}}$ is an analog of crystal momentum, and $n$ is a collective index denoting all the other quantum numbers necessary to specify the state. The corresponding eigenenergies will be continuous functions of $\theta_{\bar{\bm{x}}}$, as in conventional band theory.

For example, a single square plaquette in the ``electric limit'' $K\!\to\!0$
(the analog of the ``empty lattice approximation'' in the Bloch problem) is governed by the
Hamiltonian 
\begin{equation}
H_{g}\approx\frac{1}{2}\sum_{\bm{x},i}e_{\bm{x},i}^{2}.\label{eq:elH}
\end{equation}
There is a single site $\bar{\bm{x}}$ on the dual lattice, and we
thus drop the site index. Eigenstates of $H_{g}$ are eigenstates
of all four electric fields bordering the plaquette, but subject to
the Bloch condition (\ref{eq:bloch}) and the Gauss constraint $(\mathrm{div}e)_{\bm{x}}\ket{\Psi}\!=\!0$.
For this single-plaquette system, Eq.~(\ref{eq:monop1}) implies $\sigma_{\bar{\bm{x}}}\!=\!e_{\bm{x},\hat{x}_{1}}$
and thus $\mathcal{M}_{Q}(\bar{\bm{x}})\!=\!\exp(i2\pi Qe_{\bm{x},\hat{x}_{1}})$.
The Bloch condition (\ref{eq:bloch}) is then
\begin{equation}
e^{i2\pi Qe_{\bm{x},\hat{x}_{1}}}\ket{\Psi_{n,\theta}}=e^{iQ\theta}\ket{\Psi_{n,\theta}}.
\end{equation}
 This implies physical eigenstates of $e_{\bm{x},\hat{x}_{1}}$ (and of
$H_{g}$ in the limit $K\!\to\!0$) satisfying the Bloch condition
are restricted to those with eigenvalues 
\begin{equation}
e_{\bm{x},\hat{x}_{1}}=n+\frac{\theta}{2\pi},\qquad n\!\in\!\mathbb{Z}.
\end{equation}
The electric fields on the other links can be found using the Gauss
constraint. The physical states are loops of electric flux circling
the plaquette, with an integer level spacing. Substitution of these
values into the Hamiltonian (\ref{eq:elH}) gives the bandstructure
\begin{equation}
E_{n}(\theta)=2\left(n+\frac{\theta}{2\pi}\right)^{2}.
\end{equation}
In the minimal compactness picture, $\{\theta_{\bar{\bm{x}}}\}$ are
a set of quantum numbers specifying states in the same Hilbert space.

Finally, we observe that $[H,\c{M}_Q(\bar{\b{x}})]\!=\!0$ for the full lattice Hamiltonian $H$ with gauge fields and fermions, since the gauged fermion hopping term discussed in Sec.~\ref{sec:LGT} is invariant under local shifts of the link field $a_{\b{x},i}$ by arbitrary integer multiples of $2\pi$, including those produced by conjugation with the monopole operator (\ref{eq:monop1}). Thus the Bloch condition (\ref{eq:bloch}) applies to eigenstates of $H$ as well. As with the Bloch theorem in solid-state physics, an eigenfunctional of $H$ satisfying this
Bloch condition can be written as the product of
a ``plane wave'' and a periodic function, 
\begin{equation}
\Psi_{n,\theta}[a_{\bm{x},i}]=e^{\frac{i}{2\pi}\sum_{\bar{\bm{x}}}\theta_{\bar{\bm{x}}}f_{\bar{\bm{x}}}}\Phi_{n,\theta}[a_{\bm{x},i}],\label{eq:bltwist}
\end{equation}
 where $\Phi_{n,\theta}$ is invariant under $2\pi Q$ flux translations
on a plaquette, i.e., $\mathcal{M}_{Q}(\bar{\bm{x}})\Phi\!=\!\Phi$, and we have suppressed the dependence of $\Phi$ on fermionic coordinates, which does not play a role in this analysis. As in band theory, we can reduce the solution of the Schrödinger equation
over $a_{\bm{x},i}\!\in\!\mathbb{R}$ to that over a single ``unit
cell'' $a_{\bm{x},i}\!\in\![0,2\pi)$ by either solving the original
equation over that domain with the twisted periodic boundary conditions
(\ref{eq:bloch}), or by deriving an equation for the periodic part
$\Phi$. Defining the unitary
\begin{equation}
U_{\theta}=e^{\frac{i}{2\pi}\sum_{\bar{\bm{x}}}\theta_{\bar{\bm{x}}}f_{\bar{\bm{x}}}},
\end{equation}
we see that $\Phi_{n,\theta}$ obeys the modified Schrödinger equation
$H_{\theta}\Phi_{n,\theta}\!=\!E_n(\theta)\Phi_{n,\theta},$
where
\begin{align}
H_{\theta}&\equiv  U_{\theta}^{\dagger}HU_{\theta}\nonumber \\
&=  \frac{1}{2}\sum_{\bm{x},i}\left[e_{\bm{x},i}\!+\!\frac{1}{2\pi}\epsilon_{ij}\Delta_{j}\theta_{\bar{\bm{x}}-\hat{x}_{j}}\right]^{2}\!+\!K\sum_{\bar{\bm{x}}}(1\!-\!\cos f_{\bar{\bm{x}}}).\label{eq:Htheta}
\end{align}
Since $\theta_{\bar{\bm{x}}}$ only enters $H_\theta$ through its spatial lattice derivative, a uniform parameter $\theta_{\bar{\bm{x}}}\!=\!\theta$ has no effect in the bulk~\cite{vergeles1979,brown1997}, but will affect energetics in a system with boundary as the single square plaquette considered here.
 

The partition function for a fixed set $\{\theta_{\bar{\bm{x}}}\}$
is \citep{zinnjustin2004}
\begin{equation}
Z_{\theta}=\tr e^{-\beta H_{\theta}}=\sum_{\{Q_{\bar{\bm{x}}}\}\in\mathbb{Z}}e^{i\sum_{\bar{\bm{x}}}Q_{\bar{\bm{x}}}\theta_{\bar{\bm{x}}}}Z_{Q},
\end{equation}
 where the second equality is a Fourier decomposition, since $Z_{\theta}$
is periodic in all the $\theta_{\bar{\bm{x}}}$. The Fourier coefficients
are given by 
\begin{equation}
Z_{Q}=\int_{0}^{2\pi}D\{\theta_{\bar{\bm{x}}}\}\,e^{-i\sum_{\bar{\bm{x}}}Q_{\bar{\bm{x}}}\theta_{\bar{\bm{x}}}}Z_{\theta},
\end{equation}
where we define $D\{\theta_{\bar{\bm{x}}}\}\equiv\prod_{\bar{\bm{x}}}\frac{\dd{\theta_{\bar{\bm{x}}}}}{2\pi}$. These Fourier coefficients can be interpreted as a partition function of the original Hamiltonian
$H$ with monopole insertions as follows. Let $\mathcal{M}(Q_{\bar{\bm{x}}})$
be a product of monopole operators that inserts flux across the system
in a manner determined uniquely by the configuration function $Q_{\bar{\bm{x}}}$.
Using the completeness of flux $(\hat{f})$ eigenstates in the gauge-invariant subspace, we obtain:
\begin{align}
\tr e^{-\beta H}\mathcal{M}(Q_{\bar{\bm{x}}}) & =\int_{\mathbb{R}}D\{f_{\bar{\bm{y}}}\}\bra{\{f_{\bar{\bm{y}}}\}}e^{-\beta H}\mathcal{M}(Q_{\bar{\bm{x}}})\ket{\{f_{\bar{\bm{y}}}\}}\nonumber \\
 & =\int_{\mathbb{R}}D\{f_{\bar{\bm{y}}}\}\bra{\{f_{\bar{\bm{y}}}\}}e^{-\beta H}\ket{\{f_{\bar{\bm{y}}}+2\pi Q_{\bar{\bm{y}}}\}},
\end{align}
where $D\{f_{\bar{\bm{y}}}\}\equiv\prod_{\bar{\bm{y}}}\frac{\dd{f_{\bar{\bm{y}}}}}{2\pi}$. Inserting a complete set of eigenstates of $H$ using 
\begin{equation}
\mathbb{I}=\int_{0}^{2\pi}D\{\theta_{\bar{\b{x}}}\}\sum_{\{n_{\bar{\bm{x}}}\}}\ket{\Psi_{n,\theta}}\bra{\Psi_{n,\theta}},
\end{equation}
 we find that
\begin{align}
&\tr e^{-\beta H}\mathcal{M}(Q_{\bar{\bm{x}}})  =\int_{0}^{2\pi}D\{\theta_{\bar{\b{x}}}\}\int_{\mathbb{R}}D\{f_{\bar{\bm{y}}}\}\sum_{\{n_{\bar{\bm{x}}}\}}\nn\\
&\hspace{10mm}\phantom{=}\times\bra{\{f_{\bar{\bm{y}}}\}}e^{-\beta H}\ket{\Psi_{n,\theta}}\braket{\Psi_{n,\theta}}{\{f_{\bar{\bm{y}}}+2\pi Q_{\bar{\bm{y}}}\}}\nonumber \\
&\hspace{10mm} =\int_{0}^{2\pi}D\{\theta_{\bar{\b{x}}}\}
\sum_{\{n_{\bar{\bm{x}}}\}}e^{-\beta E_n(\theta)}e^{-i\sum_{\bar{\bm{x}}}Q_{\bar{\bm{x}}}\theta_{\bar{\bm{x}}}}
\nn\\
&\hspace{10mm}\phantom{=}\times\int_{\mathbb{R}}D\{f_{\bar{\b{y}}}\}
\Psi_{n,\theta}^{*}[f]\Psi_{n,\theta}[f]\nonumber \\
&\hspace{10mm} =\int_{0}^{2\pi}D\{\theta_{\bar{\b{x}}}\}\,e^{-i\sum_{\bar{\bm{x}}}Q_{\bar{\bm{x}}}\theta_{\bar{\bm{x}}}}\tr e^{-\beta H_{\theta}}\nonumber \\
&\hspace{10mm} =Z_{Q}.
\end{align}
In the second line, we have used the Bloch condition
satisfied by the gauge-invariant wavefunctional $\Psi_{n,\theta}[f]$
as seen in Eq. (\ref{eq:bltwist}), and the third line follows from
its normalization to unity. Therefore, the partition function for
a fixed set $\{\theta_{\bar{\bm{x}}}\}$ can be written as
\begin{align}
Z_{\theta} & =\sum_{\{Q_{\bar{\bm{x}}}\}\in\mathbb{Z}}e^{i\sum_{\bar{\bm{x}}}Q_{\bar{\bm{x}}}\theta_{\bar{\bm{x}}}}\tr e^{-\beta H}\mathcal{M}(Q_{\bar{\bm{x}}}).\label{eq:ptfn}
\end{align}
Each term of this series can be written as a path integral in a fixed
monopole configuration background. The $Q$-dependent exponential
prefactor can be absorbed into the trace by explicitly including
a $\theta$-term $i\int\theta(x)\!\star\!\dd{f(x)}$ in the action,
where $f(x)$ is now the Euclidean electromagnetic 2-form and $\star$ denotes the Hodge star.

\subsection{Forced compactness}\label{sec:forced}

In the forced compactness picture, the gauge field $a_{\bm{x},i}\!\in\!U(1)\!\cong\mathbb{R}/2\pi\mathbb{Z}$
is a rotor-valued variable. The canonically conjugate electric fields
then have a spectrum valued in $\mathbb{Z}$. In this perspective,
the various minima $f_{\bar{\bm{x}}}\!=\!2n_{\bm{x}}\pi$ for $n_{\bm{x}}\!\in\!\mathbb{Z}$
are identified as the \emph{same} state, and a flux translation $f_{\bar{\bm{x}}}\!\to\!f_{\bar{\bm{x}}}\!+\!2\pi Q$ becomes a gauge redundancy.

In this perspective, the problem is akin
to that of a quantum particle on a ring, where a translation by a
length equal to the circumference is a gauge transformation. If the
ring is suspended in a gravitational potential, then the unique classical
ground state that minimizes the potential is at the bottom of the
ring. However, in the quantum problem, there are tunneling events
(instantons) that correspond to the particle winding around the ring
an integer number of times, which involves overcoming a potential
barrier.

The exact analog in CQED$_3$ in the forced compactness picture
are monopole-instantons that cause $f_{\bar{\bm{x}}}\!\to\!f_{\bar{\bm{x}}}\!+\!2\pi Q$
on a plaquette. The monopole operator defined by Eqs.~(\ref{eq:monop})-(\ref{eq:monop1})
is thus a gauge transformation (a ``do-nothing'' operator) that connects
different labels for the same physical state. These are called \emph{large
gauge transformations, }terminology inspired by analogous concepts
in 4D Yang-Mills theory~\citep{jackiw1976,callan1976}. Large gauge transformations are distinguished
from the usual \emph{small }ones in that the former crucially utilize
the multi-valuedness of the gauge function. A small gauge transformation
is one of the form $\prod_{\bm{x}}\exp[i\phi_{\bm{x}}(\mathrm{div}e)_{\bm{x}}]$,
where $\{\phi_{\bm{x}}\}$ are single-valued gauge functions, i.e.,
all of them lie in a single branch of $\mathbb{R}/2\pi\mathbb{Z}$, for
example $[0,2\pi)$. One example of a large gauge transformation is
$\prod_{p=0}^{\infty}\exp[i0(\mathrm{div}e)_{\bm{x}+p\hat{x}_{1}}]$ which,
since $0\!\sim\!2\pi$ in $\mathbb{R}/2\pi\mathbb{Z}$, can be written
as $\prod_{p=0}^{\infty}\exp(i2\pi e_{x+p\hat{x}_{1},\hat{x}_{2}})$, which is
the same operator as (\ref{eq:monop1}), but with a displaced Dirac
string.

Physical states are required to be invariant under all gauge transformations,
small and large. This imposes $\theta_{\bar{\bm{x}}}\!=\!0$ for all
plaquettes $\bar{\bm{x}}$ in the Bloch condition (\ref{eq:bloch}).
For the particle on a ring, a background flux can be threaded through
the ring, which \emph{changes }the Hamiltonian and the Hilbert space
of the problem, as we are dealing with a physically different system.
The background flux can be unitarily removed from the Hamiltonian,
at the expense of twisting the boundary conditions on wavefunctions,
which in winding around the ring, will then gain an Aharonov-Bohm
phase. Similarly, in CQED$_3$, one can introduce a \emph{theta term}, which changes the Hamiltonian from Eq.~(\ref{eq:Hg}) to Eq.~(\ref{eq:Htheta}).
There is a macroscopic number of such possible theta terms, corresponding
to a choice $\{\theta_{\bar{\bm{x}}}\}$. Again, one can remove the
theta terms from the Hamiltonian, but at the expense of introducing
twisted boundary conditions under large gauge transformations, as
in Eq.~(\ref{eq:bloch}). The key difference with the minimal compactness
picture is that a given set $\{\theta_{\bar{\bm{x}}}\}$ labels the
entire Hilbert space of the theory, sometimes called a given \emph{theta
universe}. States with different $\{\theta_{\bar{\bm{x}}}\}$ belong
in different Hilbert spaces; conversely, states in the same Hilbert
space have the same $\{\theta_{\bar{\bm{x}}}\}$.

The expression (\ref{eq:ptfn}) for a partition function with fixed
$\{\theta_{\bar{\bm{x}}}\}$ remains valid in the forced compactness
perspective. In fact, it is the full (i.e., unrestricted) partition
function here since the entire Hilbert space is characterised by the
fixed set of parameters $\{\theta_{\bar{\bm{x}}}\}$.


\section{Zero modes of massive fermions in instanton backgrounds\label{sec:massmars}}

Having discussed instantons in the pure gauge theory, we now include fermions. As mentioned previously, the presence of fermionic ZMs in instanton backgrounds is typically associated with massless fermions in non-Abelian gauge fields~\cite{thooft1976,thooft1976a,affleck1982,callias1978,bott1978}. Marston~\citep{marston1990} considered massless Dirac fermions in an Abelian instanton background in (2+1)D and found no fermion
ZMs bound to the instanton. Motivated by the $U(1)$ parton gauge theory (\ref{eq:ptnL}), we consider here \emph{massive }Dirac
fermions in the same background and show by explicit
construction that fermion ZMs now exist. This result is in accordance with the existence of zero-energy
bound states for relativistic fermions in a (soliton) monopole background
in (3+1)D (see Ref.~\citep{yamagishi1983} and references therein).
In the soliton version of the problem, the fermion ZM is found by
a self-adjoint extension of the Dirac Hamiltonian.
Such a technique is inapplicable for the instanton version of the problem
as the Euclidean Dirac operator $\c{D}$ appearing in the action [see Eq.~(\ref{EuclL})] is not Hermitian,
nor is there any requirement for it to be. Rather, $\c{D}$ must obey reflection positivity (see Appendix~\ref{app:SAop}). Therefore, the calculation
of the ZM solution must be done anew in the context of the instanton
problem.

As Marston himself notes, the Callias index theorem for odd-dimensional
noncompact manifolds provides the number of fermion ZMs in the case
of massless fermions in the background of non-Abelian instantons~\cite{callias1978,bott1978}. This
index theorem crucially relies on (i) the existence of a Higgs field,
and (ii) on relating the index of the Dirac operator $\mathcal{D}$
to $(\dim\ker\mathcal{D}^{\dagger}\mathcal{D}\!-\!\dim\ker\mathcal{D}\mathcal{D}^{\dagger})$,
both of which fail to hold in the current setting of massive fermions
in Abelian instanton backgrounds. The reason for the failure of (ii)
might seem surprising, and is discussed in Appendix~\ref{app:SAop}. Despite the absence of a rigorous index theorem for the current problem, and as we discuss below, the ZMs we find by explicit solution can be given a topological interpretation by analogy with Hamiltonian (quasi-)zero modes associated with the Atiyah-Singer index theorem.

\subsection{Setting up in spherical coordinates}

A 3D Dirac fermion $\psi$ of charge $e$ (with sign) and mass $m$,
in an instanton background $a_{\mu}^{(g)}$ of topological charge $g$, is specified
by the Euclidean Lagrangian 
\begin{align}\label{EuclL}
\mathcal{L} & =\bar{\psi}(\slashed{\partial}-ie\slashed{a}^{(g)}+m)\psi\equiv\bar{\psi}\mathcal{D}\psi.
\end{align}
The instanton is assumed to sit at the origin, with its Dirac-monopole vector potential
defined à la Wu and Yang \citep{wu1976}. Working in spherical coordinates $(r,\theta,\varphi)$, we have $a_{r}^{(g)}\!=\!a_{\theta}^{(g)}\!=\!0$ and 
\begin{equation}
\label{eq:wuyangA}
a_{\varphi}^{(g)}=\begin{cases}
-\frac{g}{r\sin\theta}(\cos\theta-1), & \mathbf{r}\in R_{N},\\
-\frac{g}{r\sin\theta}(\cos\theta+1), & \mathbf{r}\in R_{S},
\end{cases}	
\end{equation}
where charts in the northern and southern hemispheres, $R_{N}$ and $R_{S}$,
are defined by $R_{N}\!:\!\theta\!\in\![0,\pi/2\!+\!\delta)$ and $R_{S}:\theta\!\in\!(\pi/2\!-\!\delta,\pi]$,
and a choice of $\delta\!\in\![0,\pi/2)$ defines the chart overlap
region $R_{NS}:\theta\!\in\!(\pi/2\!-\!\delta,\pi/2\!+\!\delta)$.
The Dirac matrices are Pauli matrices $(\sigma_{x},\sigma_{y},\sigma_{z})$,
with $\hat{z}$ being the Euclideanized time direction. The ZM $\psi_{0}$
of the Dirac operator $\mathcal{D}$ solves 
\begin{equation}
-i\mathcal{D}\psi_{0}\equiv(-i\slashed{\partial}-e\slashed{a}^{(g)}-im)\psi_{0}=0.
\end{equation}
We will treat this formally as a quantum mechanics problem in three
spatial dimensions, regarding $-i\partial_{j}$ as a canonical momentum
operator $p_{j}$. The ZM equation is then 
\begin{equation}
(\sigma_{j}p_{j}-e\sigma_{j}a_{j}^{(g)}-im)\psi_{0}\equiv(\sigma_{j}\pi_{j}-im)\psi_{0}=0,
\end{equation}
where the mechanical momentum $\b{\pi}\!=\!\b{p}\!-\!e\b{a}^{(g)}$. Defining $\slashed{\pi}=\b{\sigma}\cdot\b{\pi}$, we use the fact that $(\bm{\sigma}\cdot\hat{\mathbf{r}})^{2}\!=\sigma_{r}^{2}\!=\!1$
to write
\begin{align}
\slashed{\pi} & =(\bm{\bm{\sigma}}\cdot\hat{\mathbf{r}})(\bm{\bm{\sigma}}\cdot\hat{\mathbf{r}})(\b{\sigma}\cdot\b{\pi})\nonumber \\
 & =(\bm{\bm{\sigma}}\cdot\hat{\mathbf{r}})[\hat{\mathbf{r}}\cdot\bm{\pi}+i\bm{\sigma}\cdot(\hat{\mathbf{r}}\times\bm{\pi})].\label{eq:pi1}
\end{align}
The canonical angular momentum that is conserved in a monopole field
is (see Appendix~\ref{app:monmisc})
\begin{equation}
\bm{L}=\mathbf{r}\times\bm{\pi}-eg\hat{\mathbf{r}},
\end{equation}
 where $eg\!\in\!\mathbb{Z}/2$ by the Dirac quantization condition.
This can be used to rewrite Eq. (\ref{eq:pi1}) as 
\begin{equation}
\slashed{\pi}=(\bm{\bm{\sigma}}\cdot\hat{\mathbf{r}})\left(\hat{\mathbf{r}}\cdot\bm{\pi}+\frac{i}{r}\bm{\sigma}\cdot\bm{L}+\frac{ieg}{r}\bm{\sigma}\cdot\hat{\mathbf{r}}\right).
\end{equation}
Since $\bm{L}$ generates spatial rotations, which leave $r\!=\!\abs{\mathbf{r}}$
invariant, $[r,\bm{L}]\!=\!0$ and the placement of $r$ does not
matter in the above equation. Using $\bm{\pi}\!=\!\bm{p}\!-\!e\bm{a}^{(g)}$
and since $\hat{\mathbf{r}}\cdot\bm{a}^{(g)}\!=\!0$ for the Wu-Yang
potential, 
\begin{align}
\slashed{\pi} & =(\bm{\bm{\sigma}}\cdot\hat{\mathbf{r}})\left(\hat{\mathbf{r}}\cdot\bm{p}+\frac{i}{r}\bm{\sigma}\cdot\bm{L}+\frac{ieg}{r}\bm{\sigma}\cdot\hat{\mathbf{r}}\right)\nonumber \\
 & =-i(\bm{\sigma}\cdot\hat{\mathbf{r}})\left(\partial_{r}-\frac{1}{r}\bm{\sigma\cdot L}-\frac{eg}{r}\bm{\sigma}\cdot\hat{\mathbf{r}}\right).\label{eq:pi2}
\end{align}
The Dirac operator is thus
\begin{align}
\mathcal{D}&=  i(\bm{\bm{\sigma}}\!\cdot\!\hat{\mathbf{r}})\left(\hat{\mathbf{r}}\!\cdot\!\bm{p}+\frac{i}{r}\bm{\sigma}\!\cdot\!\bm{L}+\frac{ieg}{r}\bm{\sigma}\!\cdot\!\hat{\mathbf{r}}\right)+m\nonumber \\
&=  (\bm{\bm{\sigma}}\!\cdot\!\hat{\mathbf{r}})\left(\partial_{r}-\frac{1}{r}\bm{\sigma}\!\cdot\!\bm{L}-\frac{eg}{r}\bm{\bm{\sigma}}\!\cdot\!\hat{\mathbf{r}}\right)+m.\label{eq:ZMeqn}
\end{align}

\subsection{Zero modes of the Dirac operator\label{subsec:ZMdirac}}

Firstly, we note that $[\bm{J},\mathcal{D}]\!=\![\bm{J},i\slashed{\pi}]\!=\!0$,
since $\slashed{\pi}\!=\!\b{\sigma}\cdot\b{\pi}$ is a dot product that
remains invariant under simultaneous rotations of $\bm{\sigma}$ and
$\bm{\pi}$ generated by the \emph{total }angular momentum $\bm{J}=\b{L}+\frac{1}{2}\b{\sigma}$.
As a set of commuting observables, we take
\begin{equation}
[\bm{J}^{2},\mathcal{D}]=[J_{z},\mathcal{D}]=0.
\end{equation}
This means the angular part of eigenspinors of $\mathcal{D}$ are
the monopole spinor harmonics $\c{Y}^L_{eg,j,m_j}(\theta,\varphi)$ (see Appendix~\ref{app:monmisc}), which
informs the eigenspinor ansatz:
\begin{equation}
\psi_{j,m_{j}}\!=\!A_{j,m_{j}}(r)\mathcal{Y}_{eg,j,m_{j}}^{j\!+\!1/2}(\theta,\varphi)\!+\!B_{j,m_{j}}(r)\mathcal{Y}_{eg,j,m_{j}}^{j\!-\!1/2}(\theta,\varphi),
\end{equation}
for a given $(eg,j,m_{j})$. It is necessary to superpose both values
of the orbital angular momentum, $L_{\pm}\!=\!j\!\pm\!1/2$, that
give rise to a given total $j$ as $L$ is not a good quantum number.

For a monopole of the lowest magnetic charge $g\!=\pm\!1/2e$, the total
angular momentum can assume $j\!=\!0,1$. We will focus on the lowest
spherical wave $j\!=\!0$, for which the orbital angular momentum
$L_{-}\!=-\!1/2$ is excluded. To reduce notational clutter, we suppress
$(eg,j,m_{j})$ labels everywhere except on the monopole spinor harmonics, and write
the ZM ansatz as:
\begin{equation}
\psi_{0}(r,\theta,\varphi)=A_{\pm}(r)\mathcal{Y}_{\pm1/2,0,0}^{1/2}(\theta,\varphi).\label{eq:ZManz}
\end{equation}
The zero index on $\psi_{0}$ indicates that it is a ZM. The $\pm$
indices on the radial part $A_{\pm}(r)$ denote the value of $eg\!=\!\pm1/2$.
Finally, $\mathcal{Y}_{\pm1/2,0,0}^{1/2}$ stands for the $(L\!=\!1/2,eg\!=\!\pm1/2,j\!=\!0,m_{j}\!=\!0)$
monopole spinor harmonic. Since this ansatz is coincidentally also an eigenstate
of $\bm{L}^{2}$, we rewrite the Dirac operator in Eq. (\ref{eq:ZMeqn})
as:
\begin{equation}
\mathcal{D}=(\bm{\bm{\sigma}}\cdot\hat{\mathbf{r}})\left[\partial_{r}-\frac{1}{r}\left(\bm{J}^{2}\!-\!\bm{L}^{2}\!-\!\frac{3}{4}\right)-\frac{eg}{r}\bm{\bm{\sigma}}\cdot\hat{\mathbf{r}}\right]\!+\!m.\label{eq:Dop}
\end{equation}
 The action of $\mathcal{D}$ on the ZM ansatz (\ref{eq:ZManz}) is
then (for $eg\!=\!\pm1/2),$
\begin{multline}
(\bm{\bm{\sigma}}\cdot\hat{\mathbf{r}})\left[\mathcal{Y}_{\pm1/2,0,0}^{1/2}\partial_{r}A_{\pm}+\frac{3}{2r}A_{\pm}\mathcal{Y}_{\pm1/2,0,0}^{1/2}\right.\\
\left.\mp\frac{1}{2r}A_{\pm}(\bm{\bm{\sigma}}\cdot\hat{r})\mathcal{Y}_{\pm1/2,0,0}^{1/2}\right]+mA_{\pm}\mathcal{Y}_{\pm1/2,0,0}^{1/2}=0.\label{eq:Danz}
\end{multline}
Using Eq. (\ref{eq:srY}) of Appendix~\ref{app:monmisc} with $eg\!=\!\pm1/2$
yields:
\begin{equation}
(\bm{\sigma}\cdot\hat{\mathbf{r}})\mathcal{Y}_{\pm1/2,0,0}^{1/2}=\pm\mathcal{Y}_{\pm1/2,0,0}^{1/2},
\end{equation}
and therefore:
\begin{equation}
\partial_{r}A_{\pm}(r)+\left(\frac{1}{r}\pm m\right)A_{\pm}=0,
\end{equation}
 which have the obvious exponential solutions. Therefore, the ZMs
for $eg\!=\!\pm1/2$ are 
\begin{align}
\psi_{0}^{+}(r,\theta,\varphi) & =\frac{\sqrt{2m}}{r}e^{-mr}\mathcal{Y}_{1/2,0,0}^{1/2}(\theta,\varphi),\nonumber \\
\psi_{0}^{-}(r,\theta,\varphi) & =\frac{\sqrt{-2m}}{r}e^{mr}\mathcal{Y}_{-1/2,0,0}^{1/2}(\theta,\varphi),\label{eq:DZM}
\end{align}
which are normalizable for $m\!>\!0$ and $m\!<\!0$, respectively
(recall that there sits an $r^{2}$ in the Jacobian for spherical
integrations). The divergence at $r\!=\!0$ is superficial. The origin
is excluded in the problem due to the instanton there, mathematically
implemented by working in spherical coordinates. Alternatively, we
know that the instanton has a finite core due to the underlying lattice,
and so the derived form of the ZM is valid only at large distances
compared to the lattice constant. This has been discussed further
in the soliton version of the problem by Yamagishi \citep{yamagishi1983}.

\subsection{Zero modes of the adjoint Dirac operator\label{subsec:ZMadjDirac}}

Under an integration by parts, the Lagrangian changes from $\bar{\psi}\mathcal{D}\psi$
to $(\mathcal{D}^{\dagger}\bar{\psi}^{\dagger})^{\dagger}\psi$, so
the ZMs of the adjoint Dirac operator are also important. The adjoint
Dirac operator is
\begin{equation}
\mathcal{D}^{\dagger}=-\slashed{\partial}+ie\slashed{a}^{(g)}+m,
\end{equation}
where the adjoint is defined with respect to the standard inner product
on $L^{2}(\mathbb{R}^{3})$. In fact, the correct domains of $\mathcal{D}$
and $\mathcal{D}^{\dagger}$ are subsets of $L^{2}(\mathbb{R}^{3})$,
as discussed in Appendix~\ref{app:SAop}. The ZM equation
for $\mathcal{D}^{\dagger}$ is
\begin{equation}
i\mathcal{D}^{\dagger}\tilde{\psi}_{0}=(-i\slashed{\partial}-e\slashed{a}^{(g)}+im)\tilde{\psi}_{0}=0,
\end{equation}
which is the same as that of $\mathcal{D}$, but with the sign of $m$ reversed.
This implies the ZMs of $\mathcal{D}^{\dagger}$ for $eg\!=\!\pm1/2$
are
\begin{align}
\tilde{\psi}_{0}^{+}(r,\theta,\varphi) & =\frac{\sqrt{-2m}}{r}e^{mr}\mathcal{Y}_{1/2,0,0}^{1/2}(\theta,\varphi),\nonumber \\
\tilde{\psi}_{0}^{-}(r,\theta,\varphi) & =\frac{\sqrt{2m}}{r}e^{-mr}\mathcal{Y}_{-1/2,0,0}^{1/2}(\theta,\varphi),\label{eq:DdagZM}
\end{align}
 which are normalizable for $m\!<\!0$ and $m\!>\!0$, respectively.
We note that $\mathcal{D}$ and $\mathcal{D}^{\dagger}$ cannot both
possess normalizable ZMs simultaneously.

\subsection{Hamiltonian picture and quasi-zero modes}

Although the Callias index theorem does not directly apply to the problem considered here, the topological nature of the fermion ZMs explicitly found in the previous subsections can be understood by considering an approximate treatment of the same problem but in the Hamiltonian picture, following the approach of Refs.~\cite{alicea2005,alicea2005b,alicea2006,alicea2008}. In this approach, a single instanton is modeled as a static, infinitely thin $2\pi$ solenoidal flux to which the fermions respond: $\nabla\times\b{a}=2\pi\delta(\b{r})$, where $\b{r}\!=\!(x,y)$ now denotes the spatial coordinate. Consider first fermions of gauge charge $+1$ and zero mass. As is well known, for each fermion flavor, the corresponding 2D massless Dirac Hamiltonian possesses a single quasi-normalizable ``chiral'' zero-energy eigenstate $\psi_0\!=\!(u,0)^T$ with $u\!\sim\!f(x\!+\!iy)/r$, which can be understood as a manifestation of the Atiyah-Singer index theorem~\cite{jackiw1984}.

In the presence of a nonzero mass $m\!>\!0$, a ``zero-mode'' solution persists, again of the form $\psi_0\!=\!(u,0)^T$ with $u\!\sim\!f(x\!+\!iy)/r$, but its energy $E$ is shifted from zero to $E\!=\!m$. For fermions of gauge charge $-1$ and negative mass $-m$, there is also a single such quasi-zero mode per fermion flavor, now with wavefunction $\psi_0\!=\!(0,v)^T$ and $v\!\sim\!g(x\!-\!iy)/r$, but again energy $E\!=\!m$. For a $-2\pi$ flux background corresponding to an anti-instanton, the situation is reversed: fermions of gauge charge $+1$ and mass $m$ possess a quasi-zero mode $\psi_0\!=\!(0,v)^T$ with $v\!\sim\!g(x\!-\!iy)/r$, and fermions of gauge charge $-1$ and mass $-m$ possess a quasi-zero mode $\psi_0\!=\!(u,0)^T$ with $u\!\sim\!f(x\!+\!iy)/r$, both with energy $E\!=\!-m$.

The Hamiltonian quasi-zero modes discussed above are similar to those appearing in the ``zero'' mode dressing of monopole operators at the quantum critical point between a Dirac spin liquid and an antiferromagnet~\cite{dupuis2019,dupuis2019b,dupuis2021}. In the latter context, a spin-Hall mass $m\sigma_z$ appears spontaneously in the saddle-point free energy of the associated conformal field theory quantized on a sphere surrounding a monopole insertion, following the approach of Ref.~\cite{borokhov2002} to calculate the scaling dimension of monopole operators at conformal fixed points. This spin-Hall mass gives a mass of opposite sign to fermion flavors $\psi_\uparrow$ and $\psi_\downarrow$ of opposite spin, but the gauge charge is the same for both flavors. In our case, the two parton flavors $\psi_1$ and $\psi_2$ in Sec.~\ref{sec:LGT} play the role of $\psi_\uparrow$ and $\psi_\downarrow$, and the ``spin-Hall'' mass comes from the parton Chern numbers appropriate to a superfluid state~\cite{barkeshli2014}. Furthermore, the gauge charge is opposite for both flavors on account of the parton decomposition (\ref{eq:part}). Nonetheless, in both cases a single normalizable quasi-zero mode with energy $\pm m$ appears for each Dirac fermion flavor, as expected from the Atiyah-Singer index theorem.

Finally, the counting of instanton zero modes in Sec.~\ref{subsec:ZMdirac} and \ref{subsec:ZMadjDirac} is consistent with that of the Hamiltonian quasi-zero modes just discussed, if both $\c{D}$ and $\c{D}^\dag$ are considered. For an instanton ($g\!>\!0$), and for fermions of positive gauge charge ($eg\!=\!1/2$) and mass $m\!>\!0$, the Euclidean Dirac operator $\c{D}$ has a single normalizable zero mode $\psi_0^+\!\propto\!e^{-mr}/r$, where $r$ now denotes Euclidean spacetime distance from the center of the instanton [see Eq.~(\ref{eq:DZM})]. For fermions with negative gauge charge ($eg\!=\!-1/2$) and mass $m\!<\!0$, $\c{D}$ likewise possesses a single normalizable zero mode, $\psi_0^-\!\propto\!e^{mr}/r$. The adjoint Dirac operator $\c{D}^\dag$ has no zero modes in this case. For an anti-instanton ($g\!<\!0$), the situation is reversed, as in the Hamiltonian picture: $\c{D}^\dag$ now has normalizable zero modes, but $\c{D}$ has none. For $e\!>0\!$ and $m\!>\!0$, $\c{D}^\dag$ has a single zero mode $\tilde{\psi}^-_0\!\propto\!e^{-mr}/r$; for $e\!<\!0$ and $m\!<\!0$, the $\c{D}^\dag$ zero mode is $\tilde{\psi}^+_0\!\propto\!e^{mr}/r$ [Eq.~(\ref{eq:DdagZM})]. The fact that instantons (anti-instantons) are associated with zero modes of $\c{D}$ ($\c{D}^\dag$) is further discussed in Sec.~\ref{subsec:Qp} and \ref{subsec:Qm} and has important consequences for instanton-induced symmetry breaking.

\section{The 't Hooft vertex\label{sec:hooft}}

Using the explicit ZM solutions for massive 3D Dirac fermions found in Sec.~\ref{sec:massmars}, we now show that Abelian instantons in CQED$_3$ induce symmetry-breaking interactions for such massive fermions. The calculation here is analogous to 't Hooft's groundbreaking solution of the $U(1)$ problem in 4D quantum chromodynamics~\citep{thooft1976,thooft1976a}, which is well summarized in a review article~\cite{thooft1986} by the same author. In short, fermion ZMs cause transition amplitudes with nonzero instanton charge $Q\!\neq\!0$ to vanish when evaluated between states of equal fermion number. Instead, nonvanishing amplitudes occur between states of different fermion number: fermionic field insertions appearing in such amplitudes ``soak up'' the fermion ZMs appearing in Grassmann integration. Resumming these insertions in the Coulomb gas formalism produces a fermionic interaction term known as the 't Hooft vertex, whose symmetry is lower than that of the classical Lagrangian.


\subsection{Partitioning the partition function into instanton sectors}

To make the calculations less tedious, we consider only two species
of 3D Dirac fermions $\psi_{1},\psi_{2}$ with opposite gauge charges
$e_{1}\!=\!-e_{2}\!=\!e$ and zero net Chern number, so that their
masses satisfy $m_{1}\!=\!-m_{2}\!=\!m$. This corresponds to effectively
ignoring the valley $(\pm)$ subindex in the original Lagrangian (\ref{eq:ptnL}), which can be easily restored at the end of the calculation (see Sec.~\ref{sec:more}). The Lagrangian of interest is then
\begin{multline}\label{Lag1}
\mathcal{L}=\bar{\psi}_{1}(\slashed{\partial}-i\slashed{a}+m)\psi_{1}+\bar{\psi}_{2}(\slashed{\partial}+i\slashed{a}-m)\psi_{2}\\
\!+\!\frac{1}{4e^{2}}f_{\mu\nu}^{2}+i\theta(x)\epsilon_{\mu\nu\lambda}\partial_{\mu}f_{\nu\lambda},
\end{multline}
where an explicit $\theta$ term has been added to keep the discussion
general (see Sec.~\ref{sec:thetvac}). The fermion part of the action will be denoted as $S_{F}[a_{\mu}]$
when there is need to refer to it separately. The presence of a lattice
regulator permits monopole-instantons in this theory to have finite action.
The qualitative effects of those instantons on fermions are what we wish
to understand. To formulate this problem in terms of a path integral,
we use Eq.~(\ref{eq:ptfn}) to write the partition function in a fixed
$\theta$ universe as
\begin{equation}
Z=\sum_{Q\in\mathbb{Z}}\int D(\bar{\psi}_{\alpha},\psi_{\alpha})\left[Da_{\mu}\right]_{Q}e^{-S},\label{eq:ZQ}
\end{equation}
where $\left[Da_{\mu}\right]_{Q}$ indicates a restriction of the
integration over $a_{\mu}$ to configurations with total instanton
charge $Q/2e$. The sum over instanton configurations in Eq.~(\ref{eq:ptfn})
has been replaced by a sum solely over total charge $Q$, with integrations
over instanton collective coordinates subsumed in the measure $\left[Da_{\mu}\right]_{Q}$.
We will ``integrate out'' the instantons in the $Q\!\neq\!0$ sectors
and write a theory of fermions. A direct coupling between photons---quantum fluctuations of the gauge field about the classical instanton background---and fermions remains in this final theory, but can be neglected in the computation of the 't~Hooft vertex, which is a semiclassical effect~\cite{thooft1976,thooft1976a,thooft1986}.


Insofar as the pure gauge theory is concerned, the effect of an instanton of charge $Q/2e$ is captured by an insertion
\begin{equation}
e^{-\frac{\pi^{2}}{2e^{2}}Q^{2}}\int\dd^{3}x\,e^{i2\pi \sigma(x)},
\end{equation}
in the path integral~\citep{polyakov1977,polyakov1975,polyakov1987}.
Here, $\pi^{2}Q^{2}/2e^{2}$ is the action of a charge-$Q/2e$ instanton, $\sigma(x)$ is a compact
scalar called the dual photon---the continuum limit of the lattice variable $\sigma_{\bar{\b{x}}}$ in Eq.~(\ref{eq:monop1})---and the integration is over the instanton
position $x$. Likewise, the factor $\exp(i2\pi Q\sigma)$ is just the spacetime
representation of the monopole operator $\mathcal{M}_{Q}$~\citep{borokhov2002}, whose lattice form was given in Eq.~(\ref{eq:monop1}).
An attempt to dualize this theory in the same vein as the classic Polyakov
duality between the compact $U(1)$ gauge theory and the sine-Gordon theory~\citep{polyakov1975,polyakov1977,polyakov1987} leads to the path
integral
\begin{align}
Z&=\int D\sigma\,e^{-\frac{e^{2}}{2}\int\dd^{3}{x}(\partial_{\mu}\sigma)^{2}}\sum_{N=0}^{\infty}\frac{1}{N!}\nn\\
&\phantom{=}\times\prod_{k=1}^{N}\int\dd^{3}z_{k}\sum_{Q_{k}=-\infty}^{\infty}e^{-\frac{\pi^{2}}{2e^{2}}Q_{k}^{2}}
e^{iQ_k[2\pi\sigma(z_k)+\theta(z_k)]}\nn\\
&\phantom{=}\times\int D(\bar{\psi}_{\alpha},\psi_{\alpha})e^{-S_{F}[a_{\mu}^{Q_{k}}]}.\label{eq:Zmaster}
\end{align}

There are a number of results used in writing Eq.~(\ref{eq:Zmaster}), especially with fermions present. The gauge field in each
$Q$-sector in Eq.~(\ref{eq:ZQ}) has been formally decomposed as
\begin{equation}
a_{\mu}=a_{\mu}^{SW}+a_{\mu}^{Q},
\end{equation}
where $a_{\mu}^{SW}$ is the photon (analog of ``spin wave'' in the 2D XY/sine-Gordon duality~\cite{kosterlitz1974}) part whose coupling to fermions is neglected at the semiclassical level, and $a_{\mu}^{Q}$ describes an instanton configuration
of total charge $Q/2e$. The photon part has been dualized to the
Gaussian action for the compact scalar $\sigma$. In the above decomposition,
the photon part can be thought of as finite-action fluctuations (of
arbitrary size) around the fixed instanton solution. In addition,
the sum over charges $Q$, and integration over positions implicit
in the measure $\left[Da_{\mu}\right]_{Q}$ in Eq.~(\ref{eq:ZQ})
have been rewritten as sums over the number $N$ of instantons, their
charges $Q_{k}$, and their locations $z_{k}$.

An important assumption used here is the dilute gas approximation, which
gains new significance in the presence of fermions for the following
reason. Consider, for instance, $N\!=\!2$ with charges $Q_{1}/2e$
and $Q_{2}/2e$. It has been assumed~\cite{thooft1976,thooft1976a,thooft1986} that the gauge field for such
a configuration can be written as
\begin{equation}
a_{\mu}^{Q_{1}+Q_{2}}(x;z_{1},z_{2})\approx a_{\mu}^{Q_{1}}(x;z_{1})+a_{\mu}^{Q_{2}}(x;z_{2}),
\end{equation}
with $\abs{z_{1}\!-\!z_{2}}\!\gg\!1.$ This might seem plausible given
the assumption of a dilute instanton gas, but the consequence
is severe, for it implies that the Dirac action also separates, 
\begin{equation}
S_{F}[a_{\mu}^{Q_{1}+Q_{2}}]\approx S_{F}[a_{\mu}^{Q_{1}}]+S_{F}[a_{\mu}^{Q_{2}}].
\end{equation}
This allows the fermion path integral to be written inside the $N$
and $Q$ sums, as in Eq.~(\ref{eq:Zmaster}). If $Q_{1}\!=\!-Q_{2}$
so that the \emph{total} instanton charge is zero, one might expect
that there exist no normalizable fermion ZMs. However, the decomposition
of the action in the above form (which filters into a decomposition
of the Dirac operator) clearly allows ZMs. We shall nevertheless assume
such a decomposition; the error in the resulting partition function turns out be
of $\mathcal{O}(\lambda^{2})$, where
\begin{align}
\lambda=e^{-\pi^{2}/2e^{2}},
\end{align}
is the action for an instanton of lowest charge, and we consider the semiclassical limit $\lambda\!\ll\!1$. Moreover, even if there
are no strict (topologically protected) ZMs in this case, there will
likely be eigenmodes of the Dirac operator lying arbitrarily close
to zero, with splitting proportional to $\exp(-m\abs{z_{1}\!-\!z_{2}})$, since the ZM wave functions (\ref{eq:DZM}) for an isolated instanton decay exponentially away from the center of the instanton.




In what follows, we shall only consider the effects of a dilute gas
of instantons of elementary charge, thereby restricting the sum over
$Q_{k}$ in Eq.~(\ref{eq:Zmaster}) to $\pm1$. The contributions of instantons of higher topological charge are suppressed by increasing powers of $\lambda$.

\subsection{$Q\!=\!1$ instanton sector\label{subsec:Qp}}

In the $Q\!=\!1$ sector, the fermion part of the partition function
is
\begin{equation}
Z_{F}[a_{\mu}^{+}]\equiv\int D(\bar{\psi}_{\alpha},\psi_{\alpha})e^{-S_{F}[a_{\mu}^{+}]},\label{eq:Z1}
\end{equation}
where $a_{\mu}^{+}$ describes a single charge $1/2e$ instanton located
at $z_{+}$. We will show that this path integral is precisely zero.
Before the zero is revealed, the fermion functional measure needs
definition. For technical reasons discussed in Appendix~\ref{app:SAop},
the standard means of definition using the eigenfunctions of $\mathcal{D}^{\dagger}\mathcal{D}$
and $\mathcal{D}\mathcal{D}^{\dagger}$ does not work, since these
only span subspaces of $L^{2}(\mathbb{R}^{3})$ over which $\mathcal{D}^{\dagger}\mathcal{D}$
and $\mathcal{D}\mathcal{D}^{\dagger}$ are self-adjoint (not merely
Hermitian), and it so happens that the ZM lies outside these subspaces.
Instead, we shall proceed along physical lines; the effects of ZMs
are what we are interested in. The results of Sec.~\ref{sec:massmars}
indicate that the Euclidean Dirac operators,
\begin{align}
\mathcal{D}_{1} & \equiv\gamma^{\mu}\partial_{\mu}-i\gamma^{\mu}a_{\mu}^{+}+m,\nonumber \\
\mathcal{D}_{2} & \equiv\gamma^{\mu}\partial_{\mu}+i\gamma^{\mu}a_{\mu}^{+}-m,
\end{align}
each possess a normalizable ZM. We shall thus use a mode expansion
of the Fermi fields as 
\begin{align}
\psi_{1}(x) & =u_{0}(x-z_{+})\eta_{0}+\sideset{}{'}\sum_{i}u_{i}(x-z_{+})\eta_{i},\nonumber \\
\psi_{2}(x) & =v_{0}(x-z_{+})\chi_{0}+\sideset{}{'}\sum_{i}v_{i}(x-z_{+})\chi_{i},\label{eq:modexpn1}
\end{align}
where $\eta_{i},\chi_{i}$ are single-component Grassmann variables.
$u_{0}$ and $v_{0}$ are the respective ZMs of $\mathcal{D}_{1}$
and $\mathcal{D}_{2}$, calculated in Sec.~\ref{subsec:ZMdirac},
and the primed sum denotes non-ZM contributions. As discussed in Appendix~\ref{app:SAop}, we can either assume
that there exists some self-adjoint operator whose domain includes
the ZM, or we can use the eigenfunctions of a self-adjoint $\mathcal{D}_{\alpha}^{\dagger}\mathcal{D}_{\alpha}$
(which excludes the ZM) to account for non-ZM contributions and add the ZM by hand. Either way, the ZM contribution has to be accounted
for on physical grounds.

Since $\mathcal{D}_{1}^{\dagger}$ and $\mathcal{D}_{2}^{\dagger}$
are deprived of normalizable ZMs in the $Q\!=\!+1$ sector, we write
mode expansions of $\bar{\psi}_{1}\!\in\!\mathrm{Dom}(\mathcal{D}_{1}^{\dagger})$
and $\bar{\psi}_{2}\!\in\!\mathrm{Dom}(\mathcal{D}_{2}^{\dagger})$
as
\begin{align}
\bar{\psi}_{1}(x) & =\sideset{}{'}\sum_{i}\bar{u}_{i}(z_{+}-x)\bar{\eta}_{i},\quad\bar{u}_{i}(z_{+}-x)\equiv\tilde{u}_{i}^{\intercal}(x-z_{+}),\nonumber \\
\bar{\psi}_{2}(x) & =\sideset{}{'}\sum_{i}\bar{v}_{i}(z_{+}-x)\bar{\chi}_{i},\quad\bar{v}_{i}(z_{+}-x)\equiv\tilde{v}_{i}^{\intercal}(x-z_{+}),\label{eq:modexpn2}
\end{align}
where the \emph{transpose acts on spin and spatial indices}, no matter
where the $^{\intercal}$ is placed. The functional measure for fermions
can now be defined as 
\begin{align}
D(\bar{\psi}_{1,}\psi_{1},\bar{\psi}_{2},\psi_{2}) & =\dd{\eta_{0}}\dd{\chi_{0}}\sideset{}{'}\prod_{i}\dd{\bar{\eta}_{i}}\dd{\eta_{i}}\dd{\bar{\chi}_{i}}\dd{\chi_{i}}\nonumber \\
 & =\dd{\eta_{0}}\dd{\chi_{0}}D'(\bar{\eta},\eta)D'(\bar{\chi},\chi).\label{eq:Fmeasure}
\end{align}
Since the ZMs $\eta_{0},\,\chi_{0}$ do not appear in the action,
the path integral for $Z_{F}[a_{\mu}^{+}]$ is killed by the measure.
Therefore, only the sector with zero instanton charge contributes
to the full partition function $(Z)$ of the theory.

However, sectors with nonzero instanton charge will contribute to
correlation functions that can ``soak up'' the ZMs. For instance, (only)
the $Q\!=\!1$ sector contributes to the correlation function
\begin{align}
&-Z\ev{\hat{T}\psi_{1}^{\intercal}(x_{1})\psi_{2}(x_{2})}  \nn\\
&\hspace{5mm}\propto-\int\dd{\eta_{0}}\dd{\chi_{0}}D'(\bar{\eta},\eta,\bar{\chi},\chi)\sum_{i}u^{\intercal}_{i}(x_{1}-z_{+})\eta_{i}\nn\\
&\hspace{30mm}\times\sum_{j}v_{j}(x_{2}-z_{+})\chi_{j}e^{-S_{F}[a_{\mu}^{+}]}\nonumber \\
 &\hspace{5mm} =u_{0}^{\intercal}(x_{1}-z_{+})v_{0}(x_{2}-z_{+})\int D'(\bar{\eta},\eta,\bar{\chi},\chi)e^{-S_{F}[a_{\mu}^{+}]},
\end{align}
where only the fermion part of the path integral has been written.
This correlation function is non-zero provided 
\begin{equation}
\abs{x_{1}-z_{+}}\sim\abs{x_{2}-z_{+}}\lesssim1/m,
\end{equation}
where $m^{-1}$ is the width of the ZM bound to the instanton at $z_{+}$. 

\begin{figure}[t]
\begin{centering}
\includegraphics[width=0.7\columnwidth,height=0.9\columnwidth,keepaspectratio]{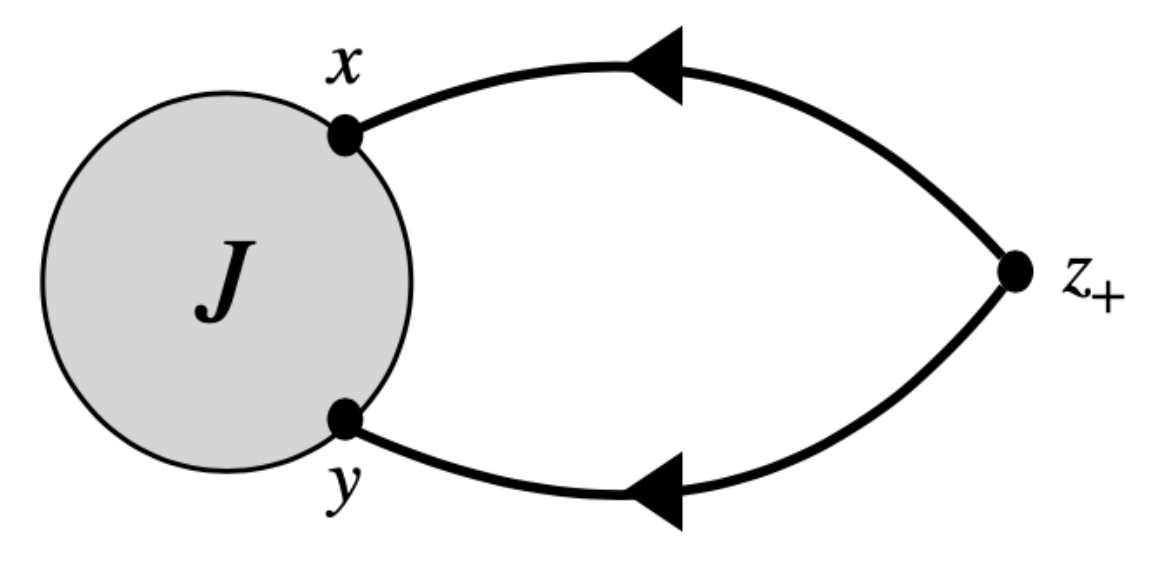}
\par\end{centering}
\centering{}\caption{Fermion pair annihilation due to the source $J(x,y)$ and instanton
at $z_{+}$.}
\label{fig:feyn}
\end{figure}

A nonzero value for such an anomalous (Gor'kov) Green's function is suggestive
of symmetry breaking. It is indeed a gauge-invariant object, since $\psi_1$ and $\psi_2$ couple to the dynamical $U(1)$ gauge field $a_\mu$ with opposite charge, but transforms nontrivially under the global $U(1)$ symmetry associated with boson number conservation in the original microscopic model (see Sec.~\ref{sec:LGT}). To investigate whether symmetry breaking indeed occurs, we
add a weak symmetry-breaking source $J$ and consider the fermion part
of the path integral in Eq. (\ref{eq:Z1}),
\begin{equation}\label{ZF+}
Z_{F}[a_{\mu}^{+},J]=\int D(\bar{\psi}_{\alpha},\psi_{\alpha})e^{-S_{F}[a_{\mu}^{+}]-\int\dd^{3}{x}\dd^{3}{y}\,\psi_{1}^{\intercal}J\psi_{2}},
\end{equation}
where the source $J(x,y)$ is generically nonlocal with some spinor
structure. This is not a valid term by itself, since it renders the
Hamiltonian non-Hermitian (or the action non reflection positive). However, the $Q\!=\!-1$
sector will provide the needed conjugate term (Sec.~\ref{subsec:Qm}). In nonlocal expressions like the source term in (\ref{ZF+}), Wilson lines should be inserted to maintain local gauge invariance. In accordance with our neglect of fermion-photon interactions at this stage, and because the final form of the 't Hooft vertex will be a local interaction, we do not write down these Wilson lines explicitly. Working to linear order
in $J$, the action can be simplified using the mode expansions in
Eqs. (\ref{eq:modexpn1})-(\ref{eq:modexpn2}) as
\begin{align}
S_{F}[a_{\mu}^{+},J] &=S_{F}[a_{\mu}^{+}]+\int\dd^{3}(x,y)\nn\\
&\phantom{=}\times\sum_{i,j}u_{i}^{\intercal}(x-z_{+})J(x,y)v_{j}(y-z_{+})\eta_{i}\chi_{j}\nonumber \\
 & =S_{F}'[a_{\mu}^{+},J]+\int\dd^{3}(x,y)\biggl[(u_{0}^{\intercal}Jv_{0})\eta_{0}\chi_{0}\nn\\
 &\phantom{=}+\sideset{}{'}\sum_{j}(u_{0}^{\intercal}Jv_{j})\eta_{0}\chi_{j}+\sideset{}{'}\sum_{i}(u_{i}^{\intercal}Jv_{0})\eta_{i}\chi_{0}\biggr],
\end{align}
where $\dd^{3}(x,y)\!\equiv\!\dd^{3}x\,\dd^{3}y$, and all the non-ZM contributions have been subsumed into $S'_{F}[a_{\mu}^{+},J]$.
Using the functional measure (\ref{eq:Fmeasure}),
\begin{align}
Z_{F}[a_{\mu}^{+},J]&=\int\dd{\eta_{0}}\dd{\chi_{0}}D'(\bar{\eta},\eta,\bar{\chi},\chi)e^{-S_{F}'[a_{\mu}^{+},J]}\nn\\
&\phantom{=}\times\biggl[1\!-\!\int\dd^{3}{(x,y)}(u_{0}^{\intercal}Jv_{0})\eta_{0}\chi_{0}\biggr]\nn\\
&\phantom{=}\times\biggl[1\!-\!\sideset{}{'}\sum_{i}\int\dd^{3}{(x,y)}(u_{0}^{\intercal}Jv_{i})\eta_{0}\chi_{i}\biggr]\nn\\
&\phantom{=}\times\biggl[1\!-\!\sideset{}{'}\sum_{i}\int\dd^{3}{(x,y)}(u_{i}^{\intercal}Jv_{0})\eta_{i}\chi_{0}\biggr],
\end{align}
the square brackets coalesce into the expression 
\begin{align}
&1\!-\!\int\dd^{3}{(x,y)}\nn\\
&\times\biggl[(u_{0}^{\intercal}Jv_{0})\eta_{0}\chi_{0}\!+\!\sideset{}{'}\sum_{i}(u_{0}^{\intercal}Jv_{i})\eta_{0}\chi_{i}\!+\!\sideset{}{'}\sum_{i}(u_{i}^{\intercal}Jv_{0})\eta_{i}\chi_{0}\biggr]\nn\\
&+\sideset{}{'}\sum_{i,j}\int\dd^{3}{(x,y)}\dd^{3}{(x',y')}(u_{0}^{\intercal}Jv_{j})(u_{i}^{\intercal}Jv_{0})\eta_{0}\chi_{j}\eta_{i}\chi_{0}.
\end{align}
All the terms except the second vanish because of the functional measure (\ref{eq:Fmeasure}); $\dd{\chi_{0}}$
kills the third, $\dd{\eta_{0}}$ the fourth, both kill the first,
and $\dd{\bar{\eta}_{i\neq0}}\dd{\bar{\chi}_{i\neq0}}$ kills the
last. Therefore, we obtain a nonvanishing result:
\begin{equation}
Z_{F}[a_{\mu}^{+},J]\!=\!\int\!\dd^{3}{x}\dd^{3}{y}\,u_{0}^{\intercal}(x\!-\!z_{+})J(x,y)v_{0}(y\!-\!z_{+})K,\label{eq:ZF1J}
\end{equation}
where $K$ denotes the contribution of nonzero modes. The order of
integration, $\dd{\eta_{0}}\dd{\chi_{0}}$, has yielded a minus sign.
The factor $K$ is independent of $J$ when working to first order
in the weak source $J$. Recalling that the ZM spinors $u_{0},v_{0}$
have radial parts of the form $\exp(-m\abs{x\!-\!z_{+}})$, similar
to the free-fermion propagator, we note that Eq. (\ref{eq:ZF1J})
resembles the structure of the Feynman diagram in Fig.~\ref{fig:feyn}.

As an ansatz for the result of integrating out instantons, we are thus
motivated to consider the path integral, 
\begin{multline}
I[J]=\int D(\bar{\psi}_{\alpha},\psi_{\alpha})e^{-S_{F}-\int\dd^{3}{x}\dd^{3}{y}\,\psi_{1}^{\intercal}(x)J(x,y)\psi_{2}(y)}\\
\times\int\dd^{3}{x_{1}}\dd^{3}{x_{2}}A\bar{\psi}_{2}(x_{2})\omega_{2}\omega_{1}^{\intercal}\bar{\psi}_{1}^{\intercal}(x_{1}),\label{eq:IJanz}
\end{multline}
where $S_{F}$ (written without a source argument) is the free Dirac
action in the \emph{absence} of instantons, $A$ is some constant to
be determined, and $\omega_{1,2}$ are spinors (possibly spacetime
dependent) to be determined. A $\bar{\psi}$ in the insertion can
pair up with a $\psi$ in the source term to give the free propagator
that we desire. $A,\omega$ are malleable quantities that must be
fixed to obtain the exact result in Eq. (\ref{eq:ZF1J}). The specific
form of the insertion is motivated in hindsight by the calculation
that follows. Taylor expanding the source exponential, we obtain:
\begin{align}
I[J] & =\int D(\bar{\psi}_{\alpha},\psi_{\alpha})e^{-S_{F}}\nn\\
&\phantom{=}\hspace{10mm}\times\left[1\!-\!\int\dd^3{(x,y)}\psi_{1}^{\intercal}(x)J(x,y)\psi_{2}(y)\right]\nn\\
&\phantom{=}\hspace{10mm}\times\int\dd^{3}{(x_{1},x_{2})}A\bar{\psi}_{2}(x_{2})\omega_{2}\omega_{1}^{\intercal}\bar{\psi}_{1}^{\intercal}(x_{1})\nonumber \\
 & =-AI[0]\int\dd^3{(x,y,x_{1},x_{2})}\nn\\
 &\phantom{=}\times\ev{\psi_{1}^{\alpha}(x)J^{\alpha\beta}(x,y)\psi_{2}^{\beta}(y)\bar{\psi}_{2}^{\gamma}(x_{2})\omega_{2}^{\gamma}\omega_{1}^{\lambda}\bar{\psi}_{1}^{\lambda}(x_{1})}_{0},
\end{align}
defining $\ev{O(\bar{\psi}_{\alpha},\psi_{\alpha})}_0\!\equiv\!\frac{1}{I[0]}\int D(\bar{\psi}_{\alpha},\psi_{\alpha})e^{-S_{F}}O(\bar{\psi}_{\alpha},\psi_{\alpha})$. The (only) effect of $I[0]$ is to restrict Wick contractions to
connected diagrams, and therefore 
\begin{align}
I[J] & =A\int\dd^3{(x,y,x_{1},x_{2})}\nn\\
&\phantom{=}\hspace{5mm}\times[G_{1}(x-x_{1})\omega_{1}]^{\intercal}J(x,y)[G_{2}(y-x_{2})\omega_{2}],\label{eq:Ij}
\end{align}
where
\begin{equation}
G(x-y)=-\ev{T_{\tau}\psi(x)\bar{\psi}(y)}_{0},
\end{equation}
 is the free, Euclidean, Dirac propagator.

Comparison of Eq. (\ref{eq:Ij}) with Eq. (\ref{eq:ZF1J}) tells us
$I[J]\!=\!Z_{F}[a_{\mu}^{+},J]$ if we make the identifications 
\begin{align}
A & =K,\nonumber \\
\int\dd^{3}{x_{1}}G_{1}(x-x_{1})\omega_{1} & =-u_{0}(x-z_{+}),\nonumber \\
\int\dd^{3}{x_{2}}G_{2}(y-x_{2})\omega_{2} & =-v_{0}(y-z_{+}).
\end{align}
The minus signs on the right-hand side are conventional, and $I[J]\!=\!Z_{F}[a_{\mu}^{+},J]$
even without them. Clearly, the second and third equalities demand
$\omega_{i}\!=\!\omega_{i}(x_{i}\!-\!z_{+})$. A shift of integration
variables $x_{i}\!\to\!x_{i}\!+\!z_{+}$ reveals that these are Fredholm
integral equations of the first kind, with solutions
\begin{align}
\omega_{1} & =-G_{1}^{-1}u_{0}=(\slashed{\partial}+m)u_{0},\nonumber \\
\omega_{2} & =-G_{2}^{-1}v_{0}=(\slashed{\partial}-m)v_{0}.\label{eq:omgs}
\end{align}
Substituting the results for $A$ and $\omega$ into Eq. (\ref{eq:IJanz})
gives 
\begin{align}
Z_{F}[a_{\mu}^{+},J]&=\int\!D(\bar{\psi}_{\alpha},\psi_{\alpha})e^{\!-\!S_{F}\!-\!\int\dd^{3}(x,y)\,\psi_{1}^{\intercal}J\psi_{2}}\int\!\dd^{3}(x_1,x_2)\nn\\
&\phantom{=}\times\bar{\psi}_{2}(x_{2})\left[K\omega_{2}(x_{2}\!-\!z_{+})\omega_{1}^{\intercal}(x_{1}\!-\!z_{+})\right]\bar{\psi}_{1}^{\intercal}(x_{1}).\label{eq:Z1fnal}
\end{align}
The single instanton of the $Q\!=\!1$ sector has been integrated out.
This path integral still evaluates to zero if the source $J$ is switched
off, since the free Dirac action $S_{F}$ conserves fermion number.
So far, all we have done is rewrite zero in different garb.

\subsection{$Q\!=\!-1$ anti-instanton sector\label{subsec:Qm}}

A similar analysis can be carried out for the $Q\!=\!-1$ sector, assuming a single charge $-1/2e$ instanton (anti-instanton) localized at $z_-$ and described by a background gauge field $a_\mu^-$. The final result can actually be written down by the requirement of reflection positivity alone, but for the sake of completeness we explicitly rederive the result here.

The anti-instanton sectors gift the adjoint Dirac operators $\mathcal{D}_{\alpha}^{\dagger}$
with ZMs, and thus contribute to anomalous Green's functions of the
form $\ev{\bar{\psi}_{2}\bar{\psi}_{1}}$. This means one has to add
a source term of the form $\bar{\psi}_{2}\tilde{J}\bar{\psi}_{1}^{\intercal}$
and consider
\begin{equation}
Z_{F}[a_{\mu}^{-},J]=\int D(\bar{\psi}_{\alpha},\psi_{\alpha})e^{-S_{F}[a_{\mu}^{-}]-\int\dd^3(x,y)\bar{\psi}_{2}\tilde{J}\bar{\psi}_{1}^{\intercal}},\label{eq:Z-1}
\end{equation}
where $\tilde{J}\!=\!\sigma_z J^\dag \sigma_z$ with $J$ being the source added in the discussion of the $Q\!=\!1$ instanton sector. This is determined by the requirement of reflection positivity of the action with the full source term ($\psi_1^\intercal J \psi_2 \!+\! \bar{\psi_2}\tilde{J}\psi_1^\intercal$).

The mode expansions are now
\begin{align}
\psi_{1}(x) & =\sideset{}{'}\sum_{i}u_{i}(x-z_{-})\eta_{i},\nonumber \\
\psi_{2}(x) & =\sideset{}{'}\sum_{i}v_{i}(x-z_{-})\chi_{i},\nonumber \\
\bar{\psi}_{1}(x) & =\bar{u}_{0}(z_{-}-x)\bar{\eta}_{0}+\sideset{}{'}\sum_{i}\bar{u}_{i}(z_{-}-x)\bar{\eta}_{i},\nonumber \\
\bar{\psi}_{2}(x) & =\bar{v}_{0}(z_{-}-x)\bar{\chi}_{0}+\sideset{}{'}\sum_{i}\bar{v}_{i}(z_{-}-x)\bar{\chi}_{i},\label{eq:modexpn2-1}
\end{align}
where $\bar{u}_{i}\!=\!\tilde{u}_{i}^{\intercal}$ and $\tilde{u}_{0}$
is the ZM of $\mathcal{D}^{\dagger}$ calculated in Sec.~\ref{subsec:ZMadjDirac}.
Inserting this into $Z_{F}[a_{\mu}^{-},J]$ gives the analog of Eq.~(\ref{eq:ZF1J}) for an anti-instanton as
\begin{align}
Z_{F}[a_{\mu}^{-},J]= & \!-\!K\!\int\!\dd^{3}{x}\dd^{3}{y}\,\bar{v}_{0}(z_{-}\!-\!x)\tilde{J}(x,y)\bar{u}_{0}^{\intercal}(z_{-}\!-\!y)\nonumber \\
= & \!-\!K\!\int\!\dd^{3}{x}\dd^{3}{y}\,\tilde{v}_{0}^{\intercal}(x\!-\!z_{-})\tilde{J}(x,y)\tilde{u}_{0}(y\!-\!z_{-}).\label{eq:ZFJ-1}
\end{align}
The Feynman-diagram interpretation of Fig.~\ref{fig:feyn} holds again but for fermion pair creation due to an anti-instanton at spacetime coordinate $z_-$. There is an extra minus sign here compared to Eq. (\ref{eq:ZF1J}),
due to the order of the measure $\dd{\eta_{0}}\dd{\chi_{0}}$. This is important
to obtain a reflection positive action at the end. The contribution from nonzero
modes, subsumed into $K$, is the same as the $Q\!=\!+1$ sector if\emph{
}the eigenfunctions of $\mathcal{D}^{\dagger}\mathcal{D}$ and $\mathcal{D}\mathcal{D}^{\dagger}$
are used to account for the non-ZM contributions in mode expansions
of fermion fields. This is because the nonzero eigenmodes of both
operators are paired with the same eigenvalues. In any case, the precise
numerical factor is unimportant here.

Similar to the analysis in the previous section, we consider with
an insertion a path integral
\begin{multline}
\tilde{I}[J]=\int D(\bar{\psi}_{\alpha},\psi_{\alpha})e^{-S_{F}-\int\dd^3{(x,y)}\bar{\psi}_{2}\tilde{J}\bar{\psi}_{1}^{\intercal}}\\
\times\int\dd^3{(x_{1},x_{2})}A\psi_{1}^{\intercal}(x_{1})\xi_{1}\xi_{2}^{\intercal}\psi_{2}(x_{2}),
\end{multline}
which can simplified as before to 
\begin{align}
\tilde{I}[J]&=-A\int\dd^3{(x,y,x_{1},x_{2})}\nn\\
&\phantom{=}\hspace{5mm}\times[\xi_{2}^{\intercal}G_{2}(x_{2}-x)]\tilde{J}(x,y)[G_{1}^{\intercal}(x_{1}-y)\xi_{1}].
\end{align}
 Demanding equality with Eq. (\ref{eq:ZFJ-1}) sets $A\!=\!K$ and
provides equations to solve for $\xi_{1,2}$:
\begin{align}
\int\dd^{3}{x_{1}}G_{1}^{\intercal}(x_{1}-y)\xi_{1}(x_{1}-z_{-}) & =\tilde{u}_{0}(y-z_{-}),\nonumber \\
\int\dd^{3}{x_{1}}G_{2}^{\intercal}(x_{2}-x)\xi_{2}(x_{2}-z_{-}) & =\tilde{v}_{0}(x-z_{-}).
\end{align}
Again, these are Fredholm integral equations of the first kind, with
solutions
\begin{align}
\xi_{1} & =-(-\slashed{\partial}+m)\tilde{u}_{0},\nonumber \\
\xi_{2} & =-(-\slashed{\partial}-m)\tilde{v}_{0}.\label{eq:xieqns}
\end{align}
The explicit forms of the ZMs, quoted here from Sec.~\ref{sec:massmars},
are 
\begin{align}
\tilde{u}_{0} & =\frac{\sqrt{2m}}{r}e^{-mr}\mathcal{Y}_{-1/2,0,0}^{1/2}(\theta,\varphi)=v_{0},\nonumber \\
\tilde{v}_{0} & =\frac{\sqrt{2m}}{r}e^{-mr}\mathcal{Y}_{1/2,0,0}^{1/2}(\theta,\varphi)=u_{0},
\end{align}
 where $u_{0},v_{0}$ are the respective ZMs of $\mathcal{D}_{1},\mathcal{D}_{2}$.
The equations (\ref{eq:xieqns}) for $\xi_{i}$ thus become 
\begin{align}
\xi_{1} & =(\slashed{\partial}-m)v_{0}=\omega_{2},\nonumber \\
\xi_{2} & =(\slashed{\partial}+m)u_{0}=\omega_{1},\label{eq:xieqns-1}
\end{align}
where $\omega_{i}$ were defined in Eq. (\ref{eq:omgs}). Therefore,
the final result for the fermion path integral in the $Q\!=\!-1$
sector is
\begin{align}
Z_{F}[a_{\mu}^{-},J]&=\int D(\bar{\psi}_{\alpha},\psi_{\alpha})e^{\!-\!S_{F}\!-\!\int\dd^3(x,y)\bar{\psi}_{2}\tilde{J}\bar{\psi}_{1}^{\intercal}}\int\!\dd^{3}(x_{1},x_{2})\nn\\
&\phantom{=}\times\psi_{1}^{\intercal}(x_{1})\left[K\omega_{2}(x_{1}\!-\!z_{-})\omega_{1}^{\intercal}(x_{2}\!-\!z_{-})\right]\psi_{2}(x_{2}).
\end{align}

\subsection{Resummation and a local Lagrangian}

Inserting the results of Sec.~\ref{subsec:Qp} and \ref{subsec:Qm} into the partition function (\ref{eq:Zmaster}) of the full theory, where only $Q=\pm 1$ instantons are kept, we obtain:
\begin{align}
&Z[J]=\int D(\bar{\psi}_{\alpha},\psi_{\alpha})D\sigma\,e^{-\int\dd^{3}x\,\mathcal{L}_0-\int\dd^{3}{(x,y)}(\psi_{1}^{\intercal}J\psi_{2}\!+\!\bar{\psi}_{2}\tilde{J}\bar{\psi}_{1}^{\intercal})}\nn\\
&\hspace{5mm}\times\sum_{N=0}^{\infty}\frac{(\lambda K)^N}{N!}\prod_{k=1}^{N}\int\dd^{3}z_{k}\int\dd^{3}x\dd^{3}y\nn\\
&\hspace{5mm}\times\left[e^{-i[2\pi\sigma(z_{k})+\theta(z_{k})]}\psi_{1}^{\intercal}(x)\omega_{2}(x\!-\!z_{k})\omega_{1}^{\intercal}(y\!-\!z_{k})\psi_{2}(y)\right.\nn\\
&\hspace{8mm}\left.+e^{i[2\pi\sigma(z_{k})+\theta(z_{k})]}\bar{\psi}_{2}(x)\omega_{2}(x\!-\!z_{k})\omega_{1}^{\intercal}(y\!-\!z_{k})\bar{\psi}_{1}^{\intercal}(y)\right],\label{eq:Zdots}
\end{align}
where
\begin{align}\label{L0}
\mathcal{L}_0=\bar{\psi}_{1}(\slashed{\partial}-i\slashed{a}+m)\psi_{1}+\bar{\psi}_{2}(\slashed{\partial}+i\slashed{a}-m)\psi_{2}+\frac{e^{2}}{2}(\partial\sigma)^{2},
\end{align}
is the Lagrangian of Eq.~(\ref{Lag1}) but absent instanton effects, with the Maxwell term dualized. We have also reinstated fermion-photon interactions to maintain explicit gauge invariance. The $k$-product in Eq.~(\ref{eq:Zdots}) just gives the $N^{\mathrm{th}}$ power of the insertion and, summing over $N$, an exponential is born. Exponentiating and \emph{then} setting
the source $J\!=\!0$ results in a nonlocal effective action 
\begin{align}
S_\text{eff}&=\int\dd^{3}{x}\,\mathcal{L}_0-\lambda K\int\dd^{3}z\dd^{3}{x}\dd^{3}{y}\nn\\
&\phantom{=}\times\Bigl[e^{-i[2\pi\sigma(z)+\theta(z)]}\psi_{1}^{\intercal}(x)\omega_{2}(x\!-\!z)\omega_{1}^{\intercal}(y\!-\!z)\psi_{2}(y)\nn\\
&\hspace{5mm}+e^{i[2\pi\sigma(z)+\theta(z)]}\bar{\psi}_{2}(x)\omega_{2}(x\!-\!z)\omega_{1}^{\intercal}(y\!-\!z)\bar{\psi}_{1}^{\intercal}(y)\Bigr].
\end{align}
Can this action be approximated by a local one? Because
the ZM wavefunctions decay exponentially in spacetime, it is reasonable to expect
so (recall Fig.~\ref{fig:feyn} and the discussion surrounding it). A change of integration variables, $x\!\to\!x\!+\!z$ and $y\!\to\!y\!+\!z$, allows the rewriting of one of the terms in the 't Hooft vertex (i.e., the instanton-induced action) as:
\begin{align}
\Delta S_\text{eff}&\equiv-\lambda K\int\dd^{3}z\dd^{3}{x}\dd^{3}{y}\,e^{-i[2\pi\sigma(z)+i\theta(z)]}\nn\\
&\hspace{15mm}\times\psi_{1}^{\intercal}(x+z)\omega_{2}(x)\omega_{1}^{\intercal}(y)\psi_{2}(y+z).
\end{align}
Since $\omega_{1}$ and $\omega_{2}$ are proportional to the radial
part $(e^{-mr}/r)$ of the ZM, the dominant contributions to the $x$
and $y$ integrals are from small neighborhoods of $x\!=\!0$ and
$y\!=\!0$. Taylor expanding the Fermi fields in powers of $x$ and
$y$ to leading (zeroth) order gives
\begin{align}\label{SeffTaylor}
\Delta S_\text{eff}&\approx-\lambda K\int\dd^{3}{z}\,e^{-i[2\pi\sigma(z)+\theta(z)]}\nn\\
&\phantom{\approx}\times\psi_{1}^{\intercal}(z)\left(\int\dd^{3}{x}\,\omega_{2}(x)\int\dd^{3}{y}\,\omega_{1}^{\intercal}(y)\right)\psi_{2}(z).
\end{align}
Using Eq. (\ref{eq:omgs}), and denoting $v_0(p)\!=\!\int\dd^3x\,e^{-ipx}v_0(x)$,
\begin{align}
\int\dd^{3}{x}\,\omega_{2}(x) & =\int\dd^{3}{x}(\slashed{\partial}-m)v_{0}(x)\nn\\
&=\lim_{p\rightarrow 0}(i\slashed p-m)v_0(p)\nn\\
&=-m\int\dd^3x\,v_0(x)\nn\\
&=-\sqrt{2}m^{3/2}\int_{0}^{\infty}\dd{r}re^{-mr}\int\dd{\Omega}\mathcal{Y}_{-1/2,0,0}^{1/2}\nn\\
&=\sqrt{\frac{2\pi}{m}}\cdot\frac{4-\sqrt{2}}{3}\begin{pmatrix}1\\
-1
\end{pmatrix},
\end{align}
and similarly, 
\begin{align}
\int\dd^{3}{x}\,\omega_{1}^{\intercal}(x) & =\int\dd^{3}{x}(\slashed{\partial}+m)u_{0}(x)\nonumber \\
 & =\sqrt{2}m^{3/2}\int_{0}^{\infty}\dd{r}re^{-mr}\int\dd{\Omega}\left(\mathcal{Y}_{1/2,0,0}^{1/2}\right)^{\intercal}\nonumber \\
 & =\sqrt{\frac{2\pi}{m}}\cdot\frac{4-\sqrt{2}}{3}\begin{pmatrix}-1 & -1\end{pmatrix},
\end{align}
we find that the quantity appearing in brackets between $\psi_{1}^{\intercal}(z)$ and $\psi_{2}(z)$ in Eq.~(\ref{SeffTaylor}) is
\begin{equation}
\int\dd^{3}{x}\,\omega_{2}(x)\int\dd^{3}{y}\,\omega_{1}^{\intercal}(y)\propto-\frac{1}{m}(\sigma_{z}+i\sigma_{y}),
\end{equation}
 where the proportionality constant is some number which shall be
subsumed into $K$. This implies the effective action is specified
by the local Lagrangian 
\begin{align}
\mathcal{L}_\text{eff}&=\mathcal{L}_0+
\frac{Ke^{-\pi^{2}/4e^{2}}}{m}\Bigl[e^{-2\pi i(\sigma+{\theta/2\pi})}\psi_{1}^{\intercal}(\sigma_{z}\!+\!i\sigma_{y})\psi_{2}\nn\\
&\hspace{10mm}+e^{2\pi i(\sigma+{\theta/2\pi})}\bar{\psi}_{2}(\sigma_{z}\!+\!i\sigma_{y})\bar{\psi}_{1}^{\intercal}\Bigr].\label{eq:Lfnal}
\end{align}
Using the transformation $\Theta$ in Appendix~\ref{app:SAop}, with additionally $\Theta(\sigma(x))=\sigma(\theta x)$ for a scalar field, one can check explicitly that the local 't Hooft vertex thus derived is reflection positive, and thus corresponds to an interaction that preserves unitarity of the underlying real-time quantum field theory.

\section{Partons and symmetry breaking}\label{sec:more}

The original parton gauge theory had $2N_f\!=\!4$ fermion flavors, described by the Lagrangian
(\ref{eq:ptnL}). In the preceding Sec.~\ref{sec:hooft}, to simplify the calculation of the 't Hooft
vertex, we retained only two fermion flavors while preserving the $U(1)$ global and gauge symmetries. It can be seen from the calculations in that section that
instantons in CQED$_3$ with $2N_{f}$ flavors of fermions with the given mass and charge assignments will induce a 't Hooft vertex with $2N_f$ fermion operators. For example, in the case of the original Lagrangian (\ref{eq:ptnL})
with four fermion flavors $\{\psi_{1\pm},\psi_{2\pm}\}$, since the mass and charge assignments are independent of the valley ($\pm$) index, one considers
the Euclidean Dirac operators:
\begin{align}
\mathcal{D}_{1} & \equiv\gamma^{\mu}\partial_{\mu}-i\gamma^{\mu}a_{\mu}+m,\nonumber \\
\mathcal{D}_{2} & \equiv\gamma^{\mu}\partial_{\mu}+i\gamma^{\mu}a_{\mu}-m.
\end{align}
The presence of a valley index for the fermion fields simply doubles the number of fermion ZMs in each instanton charge
sector. For instance, in the $Q\!=\!1$ sector, there are four ZMs, and the mode expansions of the four fermion fields now
become:
\begin{align}
\psi_{1\pm}(x) & =u_{0}(x-z_{+})\eta_{0\pm}+\sideset{}{'}\sum_{i}u_{i}(x-z_{+})\eta_{i\pm},\nonumber \\
\psi_{2\pm}(x) & =v_{0}(x-z_{+})\chi_{0\pm}+\sideset{}{'}\sum_{i}v_{i}(x-z_{+})\chi_{i\pm}.\label{eq:mdexp2}
\end{align}
This results in a path integral measure:
\begin{multline}
D(\bar{\psi}_{1\pm,}\psi_{1\pm},\bar{\psi}_{2\pm,}\psi_{2\pm})=\dd{\eta_{0}^{+}}\dd{\eta_{0}^{-}}\dd{\chi_{0}^{+}}\dd{\chi_{0}^{-}}\\
\times D'(\bar{\eta}_{\pm},\eta_{\pm})D'(\bar{\chi}_{\pm},\chi_{\pm}),
\end{multline}
where the primed measure includes contributions from nonzero modes
in the expansion (\ref{eq:mdexp2}). Therefore, to obtain a nonzero
path integral, a four-fermion insertion of the form $\psi_{1+}^{\intercal}\psi_{2+}\psi_{1-}^{\intercal}\psi_{2-}$
is required. Repeating the calculation of the 't Hooft vertex in Sec.~\ref{sec:hooft} with a four-fermion source and insertion yields an instanton-induced term (the exponentiated insertion) in the Lagrangian
of the form:
\begin{align}\label{tHooftv}
e^{-2\pi i\sigma}e^{-i\theta}\psi_{1+}^{\intercal}\gamma\psi_{2+}\psi_{1-}^{\intercal}\gamma\psi_{2-}+\mathrm{H.c.},
\end{align}
where $\gamma\!\equiv\!\sigma_z+i\sigma_y$, and we have used $\bar{\psi}\!=\!\psi^\dag\sigma_z$ to rewrite the second term as the Hermitian conjugate of the first. Because $\psi_+$ and $\psi_-$ create excitations with lattice momenta near the Dirac points $K_+$ and $K_-$, respectively, the presence of an equal number of $\psi_+$ and $\psi_-$ fields in (\ref{tHooftv}) guarantees the 't~Hooft vertex respects the microscopic translation symmetry (since $K_++K_-\!=\!0$ modulo a reciprocal lattice vector).

In the absence of instantons, (noncompact) QED$_3$ has a global topological $U(1)_\text{top}$ symmetry associated with the conservation of the topological current $j_\mu^\text{top}\!=\!\frac{i}{2\pi}\epsilon_{\mu\nu\lambda}\partial_\nu a_\lambda$. In the dual formulation, this symmetry is a shift symmetry of the dual photon $\sigma\!\rightarrow\!\sigma\!+\!\alpha$, manifest in the Lagrangian (\ref{L0}). The parton theory with noncompact gauge fluctuations thus has the global symmetry $U(1)\times U(1)_\text{top}$, where the first $U(1)$ is the boson number conservation symmetry under which $\psi_{1\pm}\!\rightarrow\!e^{i\beta}\psi_{1\pm}$ and $\psi_{2\pm}\!\rightarrow\!\psi_{2\pm}$ [recall the choice of global charge assignments in Eq.~(\ref{eq:ptnL})]. The 't Hooft vertex (\ref{tHooftv}) shows that instantons have the effect of explicitly breaking this $U(1)\times U(1)_\text{top}$ symmetry to a diagonal $U(1)$ subgroup under which
\begin{align}\label{U1diag}
\psi_{1\pm}\!\rightarrow\!e^{i\beta}\psi_{1\pm},\hspace{5mm}
\psi_{2\pm}\!\rightarrow\!\psi_{2\pm},\hspace{5mm}
\sigma\!\rightarrow\!\sigma+\frac{\beta}{\pi}\,(\text{mod } 1).
\end{align}
The latter transformation makes clear the fact that $\sigma$ is a compact scalar field of compactification radius 1. The diagonal $U(1)$ symmetry (\ref{U1diag}) is to be understood as the correct incarnation of the unique microscopic $U(1)$ boson number conservation symmetry in the low-energy parton theory with {\it compact} gauge fluctuations, i.e., where Polyakov instantons are accounted for.

Although instantons have been explicitly taken into account in the derivation of the four-fermion 't Hooft vertex (\ref{tHooftv}), the resulting effective theory is still an interacting gauge theory, and its infrared fate not altogether obvious. A natural route to confinement---our primary focus---is the instanton proliferation scenario, whereby the coefficient of the 't Hooft vertex (\ref{tHooftv}) is assumed to run to strong coupling under renormalization group flow. One then expects spontaneous breaking of the global $U(1)$ symmetry (\ref{U1diag}), with $\sigma$ acquiring an expectation value~\cite{affleck1982,unsal2008,unsal2008b}. The $\sigma$ field itself is the Goldstone mode of the broken continuous symmetry, and the microscopic hard-core boson system becomes superfluid, as discussed in Ref.~\cite{barkeshli2014}. Additionally, (\ref{tHooftv}) shows that a constant $\theta$ parameter can be given a natural interpretation as a global shift in the phase of the condensate.

However, $\langle\sigma\rangle\!\neq\!0$ only implies that the $U(1)$ symmetry is broken to a $\mathbb{Z}_2$ subgroup under which $\psi_{1\pm}\!\rightarrow\!-\psi_{1\pm}$ ($\beta\!=\!\pi$), since the 't Hooft vertex contains two $\psi_1$ fields. In terms of the original constituent bosons, this corresponds to a boson pair condensate $\langle b(\b{x})b(\b{x}')\rangle\neq 0$ without single-particle condensation, $\langle b(\b{x})\rangle\!=\!0$, which preserves an Ising symmetry $b(\b{x})\!\rightarrow\!-b(\b{x})$ (see, e.g., Refs.~\cite{bendjama2005,schmidt2006}). In terms of the fermionic partons, the order parameter $\langle\psi_{1+}^{\intercal}\gamma\psi_{2+}\psi_{1-}^{\intercal}\gamma\psi_{2-}\rangle\!\neq\!0$ is analogous to that for charge-$4e$ superconductivity~\cite{berg2009}, but without concomitant Higgsing of the $U(1)$ gauge symmetry since (\ref{tHooftv}) is manifestly gauge invariant (recall that $\psi_1$ and $\psi_2$ carry opposite gauge charge under the dynamical gauge field).

The residual global $\mathbb{Z}_2$ symmetry in such a paired superfluid can be further broken~\cite{bendjama2005,schmidt2006}, yielding a conventional superfluid phase with single-particle condensate $\langle b(\b{x})\rangle\!\neq\!0$. In the current context, this occurs if a gauge-invariant fermion bilinear condenses, $\langle \psi_1\psi_2\rangle\!\neq\!0$. The various possible spinor/valley index structures of such a bilinear (suppressed here) allow in principle for both translationally invariant condensates and spatially modulated ones, i.e., supersolid phases.

\section{Conclusion}\label{sec:concl}

In summary, we have presented a nonperturbative study of monopole-instanton effects in a (2+1)D parton gauge theory featuring Dirac fermions coupled to a compact $U(1)$ gauge field---CQED$_3$. This parton gauge theory is meant to encapsulate the universal low-energy physics of hard-core lattice bosons in the vicinity of a multicritical point separating fractionalized phases, such as boson fractional quantum Hall states, and conventional ones. While the compactness of the gauge field becomes irrelevant in fractionalized phases, which support deconfined excitations, we focused on developing an explicit understanding of the instanton dynamics that leads to confinement in conventional phases. As our first main result, we showed that in contrast to CQED$_3$ with massless fermions---an effective gauge theory describing the Dirac spin liquid---CQED$_3$ with \emph{massive} fermions supports Euclidean fermion zero modes exponentially localized on instantons. The localization length of the zero mode ``wavefunction'' is found to be inversely proportional to the fermion mass, which in hindsight elucidates the absence, first observed by Marston, of normalizable zero modes in massless CQED$_3$. While we did not prove a rigorous index theorem guaranteeing the topological stability of such Euclidean zero modes, they were found to be in one-to-one correspondence with \emph{Hamiltonian} quasi-zero modes occurring in the context of monopole operator dressing in conformal field theories associated with spin ordering transitions out of the Dirac spin liquid. In such theories, a nonzero fermion mass arises when the theory is canonically quantized on the sphere, and the resulting Hamiltonian quasi-zero modes can be understood as ``massive deformations'' of true zero modes protected by the Atiyah-Singer index theorem.

As our second main result, we combined semiclassical methods with our zero mode solutions to show by explicit derivation that instantons mediate an effective four-fermion interaction in the gauge theory, known as the 't~Hooft vertex. This effective interaction explicitly breaks a spurious $U(1)\!\times\!U(1)_\text{top}$ symmetry of the classical parton Lagrangian to a diagonal $U(1)$ subgroup, corresponding to the physical boson number conservation symmetry of the microscopic model. Under the further assumption of confinement via instanton proliferation, we found that the 't~Hooft vertex could naturally lead to two distinct superfluid phases: an ordinary single-particle condensate, but also a boson pair condensate without single-particle condensation, in which the global $U(1)$ symmetry is only broken to $\mathbb{Z}_2$.

Looking ahead, our approach based on semiclassical instanton techniques could be used to complement the Hamiltonian monopole-operator dressing approach to confinement transitions out of the Dirac spin liquid~\cite{song2019,song2020}. Song \emph{et al.} rely solely on microscopic symmetries and write down deformations of the conformal QED$_3$ Lagrangian consisting of (dressed) monopole operator/fermion composites allowed by those symmetries. Alternatively, 't~Hooft vertices containing similar physics could be explicitly derived as follows. In the two-step route to confinement advocated by Song \emph{et al.} and mentioned earlier, a fermion mass bilinear acquires an expectation value before instanton proliferation proceeds. Applying the semiclassical methods employed here after the first step, Euclidean zero modes for the resulting massive fermions could be searched for and used to derive a 't~Hooft vertex that would encapsulate the range of symmetry-breaking phases made possible by instanton proliferation. Finally, the proof of an index theorem for massive Dirac fermions in 3D Abelian instanton backgrounds would be a desirable extension of the results presented here.

\acknowledgements

We thank S. Dey, Y.-C. He, J. McGreevy, and W. Witczak-Krempa for helpful discussions. G.S. was supported by the Golden Bell Jar Graduate Scholarship in Physics. J.M. was supported by NSERC Discovery Grants Nos. RGPIN-2020-06999 and RGPAS-2020-00064; the CRC Program; CIFAR; a Government of Alberta MIF Grant; a Tri-Agency NFRF Grant (Exploration Stream); and the PIMS CRG program. 


\appendix

\section{Monopole miscellanea\label{app:monmisc}}

\subsection{Monopole harmonics\label{subsec:Yqlm}}

This appendix collates some well-known results on the theory of Dirac
monopoles and is mostly self-contained. We first consider a spinless
charge $e$ in the field of a static point monopole at the origin,
\begin{equation}
\mathbf{B}=\frac{g}{r^{2}}\hat{\mathbf{r}},
\end{equation}
described by a Wu-Yang vector potential $\bm{A}$ [see Eq.~\eqref{eq:wuyangA}]. Classically, the
spherical symmetry of the problem suggests conservation of angular
momentum. The natural guess $\bm{l}\!=\!\mathbf{r}\!\times\!(\bm{p}\!-\!e\bm{A})=\mathbf{r}\!\times\!m\bm{v}$, by minimal coupling, does not work because 
\begin{align}
\frac{\dd{\bm{l}}}{\dd{t}} & =\frac{\dd{\,}}{\dd{t}}(\mathbf{r}\times m\bm{v}),\nonumber \\
 & =\mathbf{r}\times m\ddot{\mathbf{r}},\nonumber \\
 & =\mathbf{r}\times e(\bm{v}\times\mathbf{B}),\nonumber \\
 & =\mathbf{r}\times\frac{eg}{r^{3}}(\bm{v}\times\mathbf{r}),\label{eq:ldot}\\
 & =\frac{eg}{mr^{3}}\bm{l}\times\mathbf{r}.
\end{align}
This is generically non-zero, suggesting $\bm{l}$ is not conserved.
Using a formula for the vector triple product, Eq.~(\ref{eq:ldot})
can be rewritten as \citep{shnir2005}
\begin{align}
\frac{\dd{\bm{l}}}{\dd{t}} & =\mathbf{r}\times\frac{eg}{r^{3}}(\dot{\mathbf{r}}\times\mathbf{r}),\nonumber \\
 & =\frac{eg}{r^{3}}\left[r^{2}\dot{\mathbf{r}}-(\mathbf{r}\cdot\dot{\mathbf{r}})\mathbf{r}\right],\nonumber \\
 & =\frac{eg}{r^{3}}\left[r^{2}\dot{\mathbf{r}}-\frac{1}{2}\frac{\dd{r^{2}}}{\dd{t}}\mathbf{r}\right],\nonumber \\
 & =eg\left[\frac{1}{r}\mathbf{r}-\frac{1}{r^{2}}\abs{\dot{\mathbf{r}}}\mathbf{r}\right],\nonumber \\
 & =\frac{\dd{\,}}{\dd{t}}\left(eg\frac{\mathbf{r}}{r}\right).
\end{align}
 This implies a conserved angular momentum 
\begin{equation}
\bm{L}=\mathbf{r}\times(\bm{p}-e\bm{A})-q\hat{\mathbf{r}},\qquad q\equiv eg\in\mathbb{Z}/2.\label{eq:wuyangL}
\end{equation}
 One can explicitly prove (post-quantization) that
\begin{equation}
[L_{i},L_{j}]=i\epsilon_{ijk}L_{k}.
\end{equation}

Since $[r^{2},\bm{L}]\!=\!0$, these two operators can be simultaneously
diagonalized and $\bm{L}$ can be studied for fixed $r$. Also, since
$[L_{z},\bm{L}^{2}]\!=\!0$ and $[L_{i},L_{j}]\!=\!i\epsilon_{ijk}L_{k}$,
we have the familiar
\begin{align}
\bm{L}^{2}Y_{q,L,M}(\theta,\varphi) & =L(L+1)Y_{q,L,M}(\theta,\varphi),\nonumber \\
L_{z}Y_{q,L,M}(\theta,\varphi) & =MY_{q,L,M}(\theta,\varphi).
\end{align}
The sections $Y_{q,L,M}$ are called \emph{monopole harmonics, }and
their exact form is gauge dependent, which means northern and southern
versions differ by a gauge transformation in the Wu-Yang formulation.
Only $L,M$ are quantum numbers, while $q$ is a parameter that determines
one complete set of harmonics. Just based on the $\mathfrak{su}(2)$
algebra, allowed values of $L$ must be a subset of $\{0,1/2,1,...\}$,
while $M\!\in\!\{-L,-L\!+\!1,...,L\}$. However,
\begin{align}
\bm{L}^{2} & =(\bm{l}-q\hat{\mathbf{r}})^{2}\nonumber \\
 & =\bm{l}^{2}+q^{2}-q(\bm{l}\cdot\hat{\mathbf{r}}+\hat{\mathbf{r}}\cdot\bm{l})\nonumber \\
 & =\bm{l}^{2}+q^{2}.
\end{align}
 For fixed $2q\!\in\!\mathbb{Z}$, this gives a bound on the eigenvalues
\begin{equation}
L(L+1)\geq q^{2}.
\end{equation}
The solution of the inequality above is $L\!\geq\!\abs{q}$. To prove
this, we may take $q\!\geq\!0$ without loss of generality as the
inequality is independent of $\sgn{q}$. Factorization and substitution
of $q\!=\!n/2$, where $n\!\in\!\mathbb{N}$, gives 
\begin{equation}
\left(L\!-\!\frac{-1\!+\!\sqrt{1\!+\!n^{2}}}{2}\right)\left(L\!-\!\frac{-1\!-\!\sqrt{1\!+\!n^{2}}}{2}\right)\geq0.
\end{equation}
 Both brackets must be of the same sign. Positivity of $L$ implies
\begin{align*}
L & \geq\frac{-1+\sqrt{1+n^{2}}}{2}.
\end{align*}
 $L\!\geq\!n/2\!=\!q$ satisfies this inequality, since for $n\!\geq\!0$,
\[
\frac{-1+\sqrt{1+n^{2}}}{2}\leq\frac{-1+\sqrt{1+n^{2}+2n}}{2}=\frac{n}{2},
\]
To see that this is the smallest satisfactory half-integral $L$,
note that the next smallest value of $L$ does not satisfy:
\[
\frac{n-1}{2}=\frac{-1+\sqrt{n^{2}}}{2}\leq\frac{-1+\sqrt{1+n^{2}}}{2}.
\]
We thus have the result that $L\!\geq\!\abs{q}$.

Written in spherical coordinates, Eq. (\ref{eq:wuyangL}) reads 
\begin{equation}
\bm{L}^{N/S}=-q\hat{\mathbf{r}}+\hat{\bm{\theta}}[i\csc\theta\partial_{\varphi}-q(\cot\theta\mp\csc\theta)]-\hat{\bm{\varphi}}i\partial_{\theta}.
\end{equation}
This implies
\begin{equation}
L_{z}=-i\partial_{\varphi}\mp q,
\end{equation}
for the $\hat{z}$ component, which has northern and southern eigenfunctions
of the form $\exp[i(M\!\pm\!q)\varphi]$. The requirement of a single-valued
wavefunction then mandates $(M\!\pm\!q)\!\in\!\mathbb{Z}$, which
is satisfied if $M$ is (half-)integral whenever $q$ is (half-)integral.
Together with $L\!\geq\!\abs{q}$, this determines the allowed values
of $(L,M)$ as 
\begin{equation}
L\in\{\abs{q},\abs{q}+1,...\},\quad M\in\{-L,-L+1,...,L\}.
\end{equation}

For completeness, we provide a general formula for the monopole harmonics
$Y_{q,L,M}$ in terms of the Wigner $D$-matrix. An elegant derivation
can be found in Ref.~\citep{stone1989}. In the \emph{northern hemisphere,}
\begin{equation}
Y_{q,L,M}(\theta_{N},\varphi)=\sqrt{\frac{2L+1}{4\pi}}\left[D_{M,-q}^{L}(\varphi,\theta,-\varphi)\right]^{*},
\end{equation}
where $\theta_{N}\!\in\![0,\pi).$ The southern versions (valid on
the south pole) are obtained by a gauge transformation,
\begin{equation}
Y_{q,L,M}(\theta_{S},\varphi)=e^{-i2q\varphi}Y_{q,L,M}(\theta_{N},\varphi).
\end{equation}
 The Wigner $D$-matrix is defined in terms of Euler angles $(\alpha,\beta,\gamma)$
as 
\begin{align}
D_{m',m}^{j}(\alpha,\beta,\gamma) & =\bra{jm'}e^{-i\alpha J_{z}}e^{-i\beta J_{y}}e^{-i\gamma J_{z}}\ket{jm},\nonumber \\
 & =e^{-im'\alpha}d_{m'm}^{j}(\beta)e^{-im\gamma}.
\end{align}
Using the formula above, the first two $q\!=\!1/2$ harmonics are
given by 
\begin{align}
Y_{\frac{1}{2},\frac{1}{2},\frac{1}{2}}(\theta_{N},\varphi) & =-\frac{1}{\sqrt{2\pi}}e^{i\varphi}\sin\frac{\theta}{2},\nonumber \\
Y_{\frac{1}{2},\frac{1}{2},-\frac{1}{2}}(\theta_{N},\varphi) & =\frac{1}{\sqrt{2\pi}}\cos\frac{\theta}{2},
\end{align}
in the north. Their southern versions are given by a gauge transformation
$\exp(-i\varphi)$. 

For $q\!=\!-1/2$, the first two northern harmonics are 
\begin{align}
Y_{-\frac{1}{2},\frac{1}{2},\frac{1}{2}}(\theta_{N},\varphi) & =\frac{1}{\sqrt{2\pi}}\cos\frac{\theta}{2},\nonumber \\
Y_{-\frac{1}{2},\frac{1}{2},-\frac{1}{2}}(\theta_{N},\varphi) & =\frac{1}{\sqrt{2\pi}}e^{-i\varphi}\sin\frac{\theta}{2},
\end{align}
 with their southern versions now obtained by a gauge transformation
$\exp(i\varphi)$. 

\subsection{Monopole spinor harmonics\label{subsec:spinYjm}}

We now consider a spin-1/2 particle of charge $e$ in a monopole background.
The total angular momentum for a spin-1/2 is
\begin{equation}
\bm{J}=\bm{L}+\frac{1}{2}\bm{\sigma}.
\end{equation}
The allowed eigenvalues of $\bm{J}^{2}$ and $J_{z}$, respectively
denoted $j(j\!+\!1)$ and $m_{j}$, follow from rules for addition
of angular momenta:
\begin{eqnarray}
j & \in & \{L-1/2,L+1/2\}=\left\{ \abs{q}\!-\!\frac{1}{2},\abs{q}\!+\!\frac{1}{2},...\right\} ,\nonumber \\
m_{j} & \in & \{-j,-j+1,...,j\}.
\end{eqnarray}
 The same rules also provide the (angular) eigensections $\mathcal{Y}_{q,j,m_{j}}^{L},$
called\emph{ monopole spinor harmonics},
\begin{align}
\mathcal{Y}_{q,j,m_{j}}^{j\!-\!1/2}(\theta,\varphi)= & \frac{1}{\sqrt{2j}}\begin{pmatrix}\sqrt{j\!+\!m_{j}}Y_{q,j\!-\!\frac{1}{2},m_{j}\!-\!\frac{1}{2}}\\
\sqrt{j\!-\!m_{j}}Y_{q,j\!-\!\frac{1}{2},m_{j}\!+\!\frac{1}{2}}
\end{pmatrix},\nonumber \\
\mathcal{Y}_{q,j,m_{j}}^{j\!+\!1/2}(\theta,\varphi)= & \frac{1}{\sqrt{2j\!+\!2}}\begin{pmatrix}-\sqrt{j\!-\!m_{j}\!+\!1}Y_{q,j\!+\!\frac{1}{2},m_{j}\!-\!\frac{1}{2}}\\
\sqrt{j\!+\!m_{j}\!+\!1}Y_{q,j\!+\!\frac{1}{2},m_{j}\!+\!\frac{1}{2}}
\end{pmatrix},\label{eq:spinMH}
\end{align}
where $Y_{q,L,M}$ are the monopole harmonics defined in Sec.~\ref{subsec:spinYjm}, and their coefficients are Clebsch-Gordan. For a given $q$, these
spinor harmonics are a complete, orthonormal set of 2-spinor eigensections
of $\bm{L}^{2},\bm{\sigma}^{2},\bm{J}^{2},\,J_{z}$.

For use in the main text, we also record here the action of $\bm{\sigma}\cdot\hat{\mathbf{r}}$
on $\mathcal{Y}_{q,j,m_{j}}^{L}$, which can be explicitly evaluated
using Eq.~(\ref{eq:spinMH}). Alternatively, since $\bm{\sigma}\cdot\hat{\mathbf{r}}$
commutes with $\bm{J}$, the most it can do is mix the $L\!=\!j\!\pm\!1/2$
states. A general formula is 
\begin{equation}
(\bm{\sigma}\cdot\hat{\mathbf{r}})\mathcal{Y}_{q,j,m_{j}}^{j\pm1/2}=a_{\pm}\mathcal{Y}_{q,j,m_{j}}^{j+1/2}+b_{\pm}\mathcal{Y}_{q,j,m_{j}}^{j-1/2}.\label{eq:srY}
\end{equation}
Substituting in Eq.~(\ref{eq:spinMH}) and using the known forms of
the monopole harmonics provides linear equations for the coefficients,
which turn out to be \citep{kazama1977}
\begin{alignat}{1}
a_{+}=-b_{-} & =\frac{2q}{2j+1},\nonumber \\
a_{-}=b_{+} & =-\frac{\sqrt{(2j+1)^{2}-4q^{2}}}{2j+1}.
\end{alignat}

\section{Self-adjoint operators\label{app:SAop}}

This appendix elaborates on some technical aspects of the path integrals
studied in Sec.~\ref{subsec:Qp}-\ref{subsec:Qm}, particularly
the difficulties involved in suitably defining the functional measure
and connections with index theorems. We shall first proceed along a standard route used in the physics
literature to define path integral measures. The path integral of
interest is
\begin{equation}
Z_{F}[a_{\mu}^{+}]\equiv\int D(\bar{\psi},\psi)e^{-\int\dd^{3}{x}\bar{\psi}\mathcal{D}\psi},\label{eq:Z1-1}
\end{equation}
where $a_{\mu}^{+}$ describes a single charge $1/2e$ instanton located
at $z_{+}$. Although an even number of fermion flavors are required
for this theory to make physical sense, a single flavor is sufficient
to highlight some of the general mathematical difficulties that arise
in this problem. The (massive) Euclidean Dirac operator $\mathcal{D}$
is defined as
\begin{align}
\mathcal{D} & \equiv\gamma^{\mu}\partial_{\mu}-i\gamma^{\mu}a_{\mu}^{+}+m,\nonumber \\
 &=(\bm{\bm{\sigma}}\!\cdot\!\hat{\mathbf{r}})\left[\partial_{r}-\frac{1}{r}\left(\bm{J}^{2}\!-\!\bm{L}^{2}\!-\!\frac{3}{4}\right)-\frac{q}{r}\bm{\bm{\sigma}}\cdot\hat{\mathbf{r}}\right]+m.\label{eq:appD}
\end{align}
The (naïvely taken) adjoint of the Dirac operator is
\begin{align}
\mathcal{D}^{\dagger} & =-\gamma^{\mu}\partial_{\mu}+i\gamma^{\mu}a_{\mu}^{+}+m\nonumber \\
 &=-(\bm{\bm{\sigma}}\!\cdot\!\hat{\mathbf{r}})\left[\partial_{r}-\frac{1}{r}\left(\bm{J}^{2}\!-\!\bm{L}^{2}\!-\!\frac{3}{4}\right)-\frac{q}{r}\bm{\bm{\sigma}}\cdot\hat{\mathbf{r}}\right]+m.\label{eq:appDdag}
\end{align}
The second lines of Eqs.~(\ref{eq:appD})-(\ref{eq:appDdag}) are
the results of Sec.~\ref{subsec:ZMdirac}-\ref{subsec:ZMadjDirac}.
These operators are not Hermitian, but one can consider the Hermitian
combinations $\mathcal{D}^{\dagger}\mathcal{D}$ and $\mathcal{D}\mathcal{D}^{\dagger}$
with eigenvalue equations 
\begin{align}
\mathcal{D}^{\dagger}\mathcal{D}u_{i}(x-z_{+}) & =a_{i}u_{i}(x-z_{+}),\nonumber \\
\mathcal{D}\mathcal{D}^{\dagger}\tilde{u}_{i}(x-z_{+}) & =\tilde{a}_{i}\tilde{u}_{i}(x-z_{+}),\label{eq:appDeig}
\end{align}
for an instanton located at $z_{+}$. 

We have used the term ``Hermitian operator'' to refer to what is called
a ``symmetric operator'' in the mathematical literature on unbounded
operators~\cite{reedsimon, akhiezer}. Acting on a Hilbert space, a densely defined Hermitian (or symmetric) operator
$\Lambda$ satisfies $(u,\Lambda v)\!=\!(\Lambda u,v)$, for any $u,v\!\in\!\mathrm{Dom}(\Lambda)$. To be self-adjoint,
that is for $\Lambda\!=\!\Lambda^{\dagger}$ to hold as an operator
equation, one also requires $\mathrm{Dom}(\Lambda)\!=\!\mathrm{Dom}(\Lambda^{\dagger})$, which does not follow from the Hermiticity condition for unbounded operators (such as the Dirac operator under consideration), and usually $\mathrm{Dom}(\Lambda)\!\subset\!\mathrm{Dom}(\Lambda^\dag)$.
Only self-adjoint operators have the desirable properties of possessing
a complete set of eigenfunctions and real eigenvalues. However, this
fact is typically ignored, and one proceeds to use the eigenfunctions
of $\mathcal{D}^{\dagger}\mathcal{D}$ and $\mathcal{D}\mathcal{D}^{\dagger}$
as a basis to facilitate a mode expansion of the Fermi fields in the
path integral, assuming such operators are indeed self-adjoint. This is usually harmless, but not so in current circumstances
as we shall momentarily show.\textbf{ }In any case, a loose mathematical
justification of this ignorance can be made by assuming that the Hermitian
operators above possess a unique, or a family, of self-adjoint extensions.
An arbitrary choice in this family will have the required property
of possessing a complete basis of eigenfunctions to facilitate mode
expansions.

However, a self-adjoint extension of a Hermitian operator $\Lambda$
involves imposing boundary conditions on its eigenfunctions which
effectively shrink or enlarge $\mathrm{Dom}(\Lambda)$ and $\mathrm{Dom}(\Lambda^{\dagger})$
until they are equal. The eigenfunctions of the self-adjoint extension
will form a complete basis only for the final domain $\mathrm{Dom}'(\Lambda)$.
Assuming $\mathcal{D}^{\dagger}\mathcal{D}$
or $\mathcal{D}\mathcal{D}^{\dagger}$ have been made self-adjoint, mode expansions of Fermi fields in terms of the eigenfunctions of these operators effectively assume that the space of fields
being integrated over in the path integral is the same as $\mathrm{Dom}(\mathcal{D}^{\dagger}\mathcal{D})$
or $\mathrm{Dom}(\mathcal{D}\mathcal{D}^{\dagger})$. This typically does not warrant close analysis since
one hopes (usually correctly) that all important physical effects
are accounted for in this procedure. In the present case, as we show below,
the ZM solution lies outside these domains and is thus missed if the eigenfunctions of $\mathcal{D}^{\dagger}\mathcal{D}$ or $\mathcal{D}\mathcal{D}^{\dagger}$ are used for a mode expansion or in the definition of the functional measure.

We start with the following paradox. The ZMs of $\mathcal{D}$
and $\mathcal{D}^{\dagger}$ were calculated in Sec.~\ref{subsec:ZMdirac} and \ref{subsec:ZMadjDirac}
respectively. For $g\!=\!-1/2e$, the operator $\mathcal{D}^{\dagger}$
has a normalizable ZM 
\begin{equation}
\tilde{\psi}_{0}^{-}(r,\theta,\varphi)=\frac{\sqrt{2m}}{r}e^{-mr}\mathcal{Y}_{-1/2,0,0}^{1/2}(\theta,\varphi),
\end{equation}
which implies 
\begin{equation}
\label{eq:appara1}
(\tilde{\psi}_{0}^{-},\mathcal{D}^{\dagger}\tilde{\psi}_{0}^{-})=0.
\end{equation}
However, using Eq.~(\ref{eq:appD}), one finds explicitly that $\mathcal{D}\tilde{\psi}_{0}^{-}\!=\!2m\tilde{\psi}_{0}^{-}$, which seems to imply 
\begin{equation}
\label{eq:appara2}
(\mathcal{D}\tilde{\psi}_{0}^{-},\tilde{\psi}_{0}^{-})=2m\norm{\tilde{\psi}_{0}^{-}}^2\neq0.
\end{equation}
The implied consequence is that $(\tilde{\psi}_{0}^{-},\mathcal{\mathcal{D}^{\dagger}}\tilde{\psi}_{0}^{-})\!\neq\!(\mathcal{D}\tilde{\psi}_{0}^{-},\tilde{\psi}_{0}^{-})$. Before a resolution of this is pointed out, we note that
$\mathcal{D}\tilde{\psi}_{0}^{-}\!=\!2m\tilde{\psi}_{0}^{-}$ implies
then that $\mathcal{D}^{\dagger}\mathcal{D}\tilde{\psi}_{0}^{-}\!=\!0$
so that $\dim\ker\mathcal{D}^{\dagger}\mathcal{D}\!-\!\dim\ker\mathcal{D}\mathcal{D}^{\dagger}\!=\!0$.
It is the latter that is typically calculated as $\mathrm{index}(\mathcal{D})$
in typical proofs of index theorems used in physics. This highlights
the difficulty in producing an index theorem for the current scenario,
and also in defining the functional measure of the path integral using
eigenfunctions of $\mathcal{D^{\dagger}D}$ or $\mathcal{D}\mathcal{D}^{\dagger}$.

The resolution of the paradox lies in a careful examination of the
domains of the operators $\mathcal{D}^{\dagger}$ and $\mathcal{D}$,
which turn out to be subspaces of square integrable functions. In the subspace of spinors $\psi$ with fixed angular part $\mathcal{Y}_{1/2,0,0}^{1/2}(\theta,\varphi)$, and for $eg\!=\!1/2$ (i.e., in the $Q\!=\!1$ instanton sector discussed in Sec.~\ref{subsec:Qp}), the action becomes:
\begin{equation}
S=\int\dd^3x\,\bar{\psi}(ip_r+m)\psi,\label{eq:toyact}
\end{equation}
where one defines a ``radial momentum operator''~\cite{paz2002},
\begin{align}
p_{r}  =-\frac{i}{2}(\hat{\mathbf{r}}\cdot\nabla+\nabla\cdot\hat{\mathbf{r}})
 =-i\left(\partial_ r+\frac{1}{r}\right).\label{eq:radmom}
\end{align}
The adjoint Dirac operator $\mathcal{D}^\dag$ in Eq.~\eqref{eq:appDdag} has been naïvely derived from the form of $\mathcal{D}$ by essentially assuming $p_r$ is Hermitian on $\mathrm{Dom}(\mathcal{D})$. However, this Hermiticity condition is violated on the ZM,
\begin{equation}
\psi_{0}(r)\propto\frac{1}{r}e^{-mr},\label{eq:appZM}
\end{equation}
of the operator $(ip_{r}\!+\!m)$ appearing in the action (\ref{eq:toyact}), for $(p_r \psi_0,\psi_0)\!\neq\!(\psi_0,p_r \psi_0)$. This is the reason for the paradoxical equations~\eqref{eq:appara1}-\eqref{eq:appara2}. However, to restrict the path integral over $\psi$ to a function space on which $p_r$ \emph{is} Hermitian is to exclude ZMs and their associated physics. To determine the space of fields ($\psi$ and $\bar{\psi}$) that one should integrate over, we use the necessary condition that the Minkowski action must be real-valued.

The reality of the Minkowski action in a unitary quantum field theory translates to Osterwalder-Schrader or reflection positivity of the corresponding Euclidean action, which is invariance $\Theta(S)=S$ under a form of complex conjugation followed by Euclidean time reversal~\cite{osterwalder1973}. This transformation acts on fermions as an involution of the Grassmann algebra~\cite{wetterich2011}, which for our particular choice of Dirac matrices can be chosen as
\begin{align}
\Theta(\psi_\alpha(x))&=\sigma_z^{\alpha\beta}\bar{\psi}_\beta(\theta x),\\
\Theta(\bar{\psi}_\alpha(x))&=\sigma_z^{\alpha\beta}\psi_\beta(\theta x),
\end{align}
where $\theta$ flips the sign of the time ($z$) coordinate. Additionally, $\Theta$ complex conjugates $c$-numbers and reverses the order of Grassmann variables, e.g., $\Theta(\psi_\alpha\psi_\beta\psi_\gamma)\!=\!\Theta(\psi_\gamma)\Theta(\psi_\beta)\Theta(\psi_\alpha)$. Gauge fields transform as $\Theta(a_0(x))\!=\!-a_0(\theta x)$ and $\Theta(a_i(x))\!=\!a_i(\theta x)$~\cite{simmons-duffin2016}. One can show that under $\Theta$, Eq.~\eqref{eq:toyact} transforms as:
\begin{align}
\Theta(S) & = S + \int\dd{\Omega}\left.(-r^{2}\bar{\psi}\psi)\right|_{r=0}^{r=\infty},
\end{align}
where the boundary term follows from an integration by parts. However, for reflection positivity
to hold, the boundary term is required to vanish. 

The upper limit of the boundary term vanishes if we require all fields to be
square integrable. Say $\psi\!\sim\!r^{-\beta}\chi$ as $r\!\to\!0$, where $\chi$ is
a Grassmann number without $r$ dependence. Then square integrability
requires $\lim_{r\!\to\!0}r^{3\!-\!2\beta}$ to exist, which
means $\beta\!<\!3/2$. Since square integrable functions are therefore
at most as singular as $r^{-3/2\!+\!\epsilon}$, where $\epsilon\!>\!0$,
the lower limit of the boundary term is at most as singular as 
\begin{equation}
\lim_{r\to0}\frac{r^{2}}{r^{3/2-\epsilon}\:r^{3/2-\delta}},
\end{equation}
 where $\delta\!>\!0$. The existence of this limit requires $\delta\!+\!\epsilon\!>\!1$.
The choice $\epsilon\!=\!\delta\!>\!1/2$ restricts both field integrations (over $\psi$ and $\bar{\psi}$) to the subset of square integrable functions that are less singular than $1/r$ at the origin. As discussed earlier, this excludes the ZM from both path integrals, over $\psi$ and $\bar{\psi}$. To remedy this, we may set $\epsilon\!=\!0$ and $\delta\!>\!1$, so that functions that behave as $1/r$ as $r\!\to\!0$ (such as the ZM $\psi_{0}$) are included in the path integration over $\psi$, but not in that over $\bar{\psi}$ to maintain reflection positivity of the Euclidean action. The problem now reduces to finding self-adjoint operators with domains as these new subspaces of $L^{2}(\mathbb{R}^{3})$, so that the path integral measure can be adequately defined. We will simply assume such operators, with eigenfunctions $\{\psi_i(r,\theta,\varphi)\}$ and $\{\tilde{\psi}_i(r,\theta,\varphi)\}$, exist and will expand the Fermi fields as 
\begin{align}
\psi &= \psi_0(r,\theta,\varphi)\eta_{0}+\sideset{}{'}\sum_{i}\psi_{i}(r,\theta,\varphi)\eta_{i},\nonumber \\
\bar{\psi} &= \sideset{}{'}\sum_{i}\tilde{\psi}_{i}(r,\theta,\varphi)\bar{\eta}_{i},
\end{align}
where the primed sum includes non-ZM contributions and $\bar{\eta}_i,\,\eta_{i}$ are independent Grassmann variables.

For $eg\!=\!-1/2$, i.e., in the anti-instanton sector $Q\!=\!-1$, the situation is reversed. In a subspace of spinors with fixed angular part $\mathcal{Y}_{-1/2,0,0}^{1/2}(\theta,\varphi)$, the action is
\begin{equation}
S=\int\dd^3x\,\bar{\psi}(-ip_r+m)\psi.\label{eq:toyact2}
\end{equation}
The operator $(-ip_r+m)$ does not have normalizable ZMs. However, integrating by parts, we obtain
\begin{equation}
S=\int\dd^3x\,[(ip_r+m)\bar{\psi}]\psi +\int\dd{\Omega}\left.(-r^{2}\bar{\psi}\psi)\right|_{r=0}^{r=\infty},\label{eq:toyact3}
\end{equation}
and the operator $(ip_r\!+\!m)$ does have the ZM~\eqref{eq:appZM}. Contrary to the $Q\!=\!1$ sector, we now include functions with the limiting behavior of the ZM in the path integral over $\bar{\psi}$, but exclude them from that over $\psi$ so that the boundary term in Eq.~\eqref{eq:toyact3} vanishes and reflection positivity is maintained. This implies mode expansions of the form
\begin{align}
\psi &= \sideset{}{'}\sum_{i}\tilde{\psi}_{i}(r,\theta,\varphi)\eta_{i},\nonumber\\
\bar{\psi} &= \tilde{\psi}_0(r,\theta,\varphi)\bar{\eta}_{0}+\sideset{}{'}\sum_{i}\tilde{\psi}_{i}(r,\theta,\varphi)\bar{\eta}_{i}.
\end{align}

\bibliographystyle{apsrev4-1}
\bibliography{cqed}

\end{document}